\documentclass[aps,prb,reprint,amsmath,amssymb,showpacs,superscriptaddress,longbibliography]{revtex4-2}
\usepackage[breaklinks=true,colorlinks,citecolor=blue,linkcolor=blue,urlcolor=blue]{hyperref}
\usepackage{graphicx}% Include figure files
\usepackage{xcolor}
\usepackage{subfigure}
\usepackage{hyperref,hypcap}
\usepackage{braket}
\usepackage{amsmath}
\usepackage{commath}
\usepackage{mathtools}
\usepackage{chemformula}
\usepackage{natbib}
\usepackage{tabularx}
\usepackage{orcidlink}
\usepackage{amsmath,xcolor}
\usepackage{physics}
\usepackage{bm}
\usepackage{bbold}
\usepackage{dsfont}
\usepackage{derivative}
\usepackage{tikz}
\usepackage{comment}
\begin{document}
	\title{Length--Velocity Gauge Equivalence of Quantum Geometric Nonlinear Conductivity in the Adiabatic DC Limit}
	\author{Shakeel Ahmad,\orcidlink{0009-0000-4994-9717}} 
	%\affiliation{Department of Physics,	University of Alabama at Birmingham, Birmingham, AL 35294, USA}
	\affiliation{Department of Physics,
		\textcolor{blue}{\href{https://www.uab.edu/home/}
			{University of Alabama at Birmingham}},
		Birmingham, Alabama 35294, USA}
	
	\author{Fei Xue\,\orcidlink{0000-0002-1737-2332}} 
	%\affiliation{Department of Physics,	University of Alabama at Birmingham, Birmingham, AL 35294, USA}
	\affiliation{Department of Physics,
		\textcolor{blue}{\href{https://www.uab.edu/home/}
			{University of Alabama at Birmingham}},
		Birmingham, Alabama 35294, USA}
	\date{\today}
	\begin{abstract}
		Nonlinear transport has emerged as a sensitive probe of quantum geometry beyond the Berry-curvature physics of linear response. However, the intrinsic second-order dc response remains conceptually subtle: Different quantum and semiclassical formulations can appear to give different static limits, with different assignments of Fermi sea and Fermi surface contributions. Here we resolve this ambiguity by developing a gauge-consistent density-matrix theory of intrinsic nonlinear conductivity in both the length gauge, where the electric field couples through the position operator, and the velocity gauge, where it enters through the vector potential. We show that the two gauges give the same
		adiabatic dc response when the same retarded continuation is used for all
		external frequencies and when the velocity gauge current includes all
		field-dependent vertices. The apparent Fermi sea terms cancel in the full
		expression, leaving a Fermi surface quantum geometric contribution determined
		by the band-normalized quantum metric. This result implies that a fully gapped
		insulator has no residual dc nonlinear Hall current in the adiabatic clean
		limit. The reactive part of the Fermi surface term agrees with the original semiclassical
		Berry-connection-polarizability response, while the dissipative Ohmic sector
		requires a more careful treatment of relaxation and impurity scattering. Our
		work establishes the length-velocity gauge equivalence for adiabatic dc quantum geometric nonlinear response and provides a foundation for using nonlinear
		transport to probe magnetic quantum geometry, especially in
		\(\mathcal{PT}\)-symmetric antiferromagnets.
	\end{abstract}
	\maketitle
	
	\section{Introduction}
	Electronic transport provides a powerful way to probe the geometric structure
	of Bloch wave functions. A quintessential example is the anomalous
	Hall effect, whose intrinsic contribution is governed by the Berry curvature
	of occupied Bloch states~\cite{Niu2010,Nagaosa2010}. In two-dimensional
	magnetic insulators, the corresponding Hall conductance can become quantized
	and is tied to a topological Chern number, giving rise to the quantum anomalous
	Hall effect~\cite{Review_QAHE}.
	Nonlinear transport extends this geometric viewpoint beyond linear response.
	In noncentrosymmetric systems, the second-order current can encode geometric
	quantities such as the Berry-curvature dipole~\cite{Sodeman2015,Xu2018,Ma2019} and band-normalized quantum metric dipole, which are closely related to
	Berry-connection polarizability in semiclassical wave-packet descriptions~\cite{Gao2014,Xiao2021,Yang2021,Gao2023_Experiment,Wang2023_Experiment}. These nonlinear responses therefore provide probes of band geometry beyond
	what is accessible from linear response alone~\cite{Mehraeen2024, Mehraeen2025,He2025,He2026}.
	
	In addition to its geometric significance, the linear anomalous Hall effect also provides an
	efficient transport probe of magnetic order.
	It is widely used to detect and track magnetization in ferromagnets~\cite{Nagaosa2010}, 
	and large anomalous Hall signals have been observed in noncollinear antiferromagnets such as
	Mn$_3$Sn and Mn$_3$Ge despite their small net magnetization~\cite{Chen2014,Kubler_2014,Nakatsuji2015}. More recently, anomalous Hall responses have also been discussed as symmetry-dependent transport fingerprints of altermagnetic order~\cite{Smejkal2020,Smejkal2022}. 
	However, this linear-response diagnostic fails in magnetic systems where 
	symmetry forbids the anomalous Hall effect. A particularly important example
	is a \(\mathcal{PT}\)-symmetric antiferromagnet: although time reversal is
	broken, the combined \(\mathcal{PT}\) symmetry forces the Berry curvature and
	hence the intrinsic anomalous Hall conductivity to vanish. This motivates
	nonlinear transport as a leading-order probe of magnetic order and quantum
	geometry in such systems~\cite{Godinho2018,Xiao2021,Yang2021,Gao2023_Experiment,Wang2023_Experiment}.

	The quantum metric contribution to nonlinear transport was first derived by
	Gao et al. using semiclassical wave-packet theory~\cite{Gao2014},
	which has provided a powerful framework for understanding linear
	transport phenomena~\cite{Niu2010}. In addition, nonlinear response can be
	formulated using fully quantum approaches, such as Kubo or density-matrix
	perturbation theory. At linear order, the semiclassical and quantum
	formulations give the same intrinsic geometric responses~\cite{Sinitsyn_2008,Nagaosa2010,Niu2010}. At second-order,
	however, their relation is more subtle. Recent quantum kinetic and Kubo
	calculations have obtained quantum-metric-induced nonlinear conductivities in
	forms that differ from the original semiclassical result
	~\cite{Das2023,Su2024,Yan2023,Kaplan2023,Kaplan2024}. In particular, some of
	these works identify an intrinsic quantum metric contribution to the dc
	conductivity that is dissipative rather than reactive, and is therefore not
	captured by the original wave-packet anomalous-velocity formulation. More
	recently, Refs.~\cite{Xiao2025,Jia2024} argued that this contribution can be recovered
	within a semiclassical framework once the field-induced correction to the
	wave-packet energy~\cite{Xiao2019} is included in addition to the anomalous velocity.
	
	A further subtlety arises even within fully quantum approaches. The electric
	field may be treated in the length gauge or in the velocity gauge. The
	equivalence of these two gauges is well established in optical and nonlinear
	optical response, provided a complete basis is used and all gauge-required
	terms are retained~\cite{Taghizadeh2017,Passos2018,Taghizadeh2018,Moore2019}. However,
	the dc adiabatic limit introduces an additional subtlety. In the length gauge,
	the field dependence resides in the density matrix, whereas in the velocity
	gauge it is distributed between the density matrix and the field-dependent current operator. 
	As a result, length gauge calculations~\cite{Das2023,Su2024,Jia2024} and velocity gauge
	calculations~\cite{Yan2023,Kaplan2023,Kaplan2024} can appear to produce
	different Fermi sea and Fermi surface decompositions unless the same
	adiabatic prescription and all current-operator terms are retained. This ambiguity in the static decomposition has been summarized recently in
	Ref.~\cite{Qiang2025}.

	In this work, we resolve this gauge ambiguity by developing a unified
	density-matrix perturbation theory for the intrinsic second-order static
	response in both the length and velocity gauges. We show explicitly that the
	two gauges give the same adiabatic dc result once the same frequency
	prescription is used and all field-dependent current operators are retained in
	the velocity gauge. In particular, all apparent Fermi sea contributions cancel in the full
	gauge-consistent response, leaving a Fermi surface quantum geometric term.
	Consequently, the clean adiabatic dc nonlinear Hall response vanishes in a
	fully gapped insulator when the chemical potential lies in the gap. 
	The resulting Fermi surface expression has the same quantum metric structure as the
	length gauge quantum calculations of Refs.~\cite{Das2023,Su2024}. 
	Our result also clarifies the relation to semiclassical descriptions. The
	reactive quantum metric part is consistent with the corresponding
	semiclassical response, whereas the dissipative contribution 
	takes a different form from the wave-packet-energy
	semiclassical result of Ref.~\cite{Jia2024}. We return to this comparison in detail in
	the Discussion.
	
	The remainder of the paper is organized as follows. In
	Sec.~\ref{sec:length_gauge}, we derive the second-order density-matrix response
	in the length gauge and separate the interband and intraband position operator contributions. 
	We then take the static adiabatic limit and show
	how the apparent Fermi sea terms cancel, leaving the gauge-consistent quantum-metric-induced Fermi surface response. In Sec.~\ref{sec:velocity_gauge}, we
	repeat the calculation in the velocity gauge. There we emphasize the role of
	the field-dependent current operators and show that the velocity gauge result
	reduces to the same static response obtained in the length gauge. In
	Sec.~\ref{sec:discussion}, we compare our result with semiclassical
	wave-packet descriptions and recent quantum-kinetic formulations, and discuss
	the implications for interpreting reactive and dissipative quantum geometric
	nonlinear currents. 
	Technical details are provided in the Appendixes.
	Appendixes~\ref{appendix:rho_length}--\ref{appendix:conductivity_LG} present
	the length gauge density-matrix derivation, the required position-operator
	identities, and the reduction of the static response into reactive and
	dissipative parts. 
	Appendixes~\ref{appendix:rho_vg}--\ref{app:conductivity_velocity_gauge} give
	the corresponding velocity gauge density matrix, the matrix-element identities
	for \(w^{ab}\) and \(u^{abc}\), and the velocity gauge conductivity, showing
	explicitly how its static adiabatic limit reduces to the length gauge result.
	
	\section{Quantum Kinematics in the Length Gauge}\label{sec:length_gauge}
	We begin from the quantum Liouville equation for the single-particle density matrix $\rho$ in the presence of an external electric field~\cite{Culcer2017,Das2023,Nandy2019,Jia2024},
	\begin{equation}
		\dv{\rho}{t}=-\frac{i}{\hbar}[H_0+H_E(t),\rho],
		\label{eq:Liouville_full}
	\end{equation}
	where $H_0$ is the unperturbed crystal Hamiltonian and $H_E(t)$ describes the coupling to the electric field. Throughout this section we work in the clean limit and therefore do not include a phenomenological collision integral or relaxation-time term in the kinetic equation. The low-frequency response is instead defined by the usual adiabatic prescription introduced below.
	
	In the length gauge, the electric-field coupling is
	\begin{equation}
		H_{\mathrm E}(t)=-q\,E^a(t)x^a,
		\label{eq:HE_length}
	\end{equation}
	where $q$ is the carrier charge ($q=-e$ for electrons). 
	
	We write the electric field in Fourier form as
	\begin{equation}
		E^a(t)=\int \frac{d\omega}{2\pi}\,E^a(\omega)e^{-i\omega t},
		\qquad
		E^a(-\omega)=E^a(\omega)^*,
		\label{eq:E_fourier}
	\end{equation}
	where the second relation enforces reality of the physical electric field. Unless otherwise specified, all frequency integrals run from $-\infty$ to $+\infty$.
	
	We solve Eq.~\eqref{eq:Liouville_full} iteratively in powers of the perturbation $H_E(t)$ by expanding $\rho=\rho^{(0)}+\rho^{(1)}+\rho^{(2)}+\cdots$.
	Matching terms order by order gives
	\begin{equation}
		\dv{\rho^{(N)}}{t}+\frac{i}{\hbar}[H_0,\rho^{(N)}]
		=
		-\frac{i}{\hbar}[H_E,\rho^{(N-1)}].
		\label{eq:rhoN_eom}
	\end{equation}
	
	The first- and second-order corrections are correspondingly expanded as
	\begin{equation}
		\rho^{(1)}(t)=\int \frac{d\omega}{2\pi}\,\rho^{(1)}(\omega)e^{-i\omega t},
		\label{eq:rho1_fourier}
	\end{equation}
	and
	\begin{equation}
		\rho^{(2)}(t)=\int \frac{d\omega_1}{2\pi}\int \frac{d\omega_2}{2\pi}\,
		\rho^{(2)}(\omega_1,\omega_2)e^{-i(\omega_1+\omega_2)t}.
		\label{eq:rho2_fourier}
	\end{equation}
	When convenient, we denote the second-order component by $\rho^{(2)}(\Omega;\omega_1,\omega_2)$, where the output frequency satisfies $\Omega=\omega_1+\omega_2$. 
	Using the Fourier expansions in Eqs.~\eqref{eq:E_fourier}--\eqref{eq:rho2_fourier}, one may further project Eq.~\eqref{eq:rhoN_eom} onto fixed output frequency. In particular, the first-order response satisfies
	\begin{equation}
		-i\omega\,\rho^{(1)}(\omega)+\frac{i}{\hbar}[H_0,\rho^{(1)}(\omega)]
		=
		-\frac{i}{\hbar}[H_{\mathrm E}(\omega),\rho^{(0)}],
		\label{eq:rho1_freq_eom}
	\end{equation}
	while the second-order component at output frequency $\Omega=\omega_1+\omega_2$ obeys
	\begin{equation}
		\begin{split}
			&-i\Omega\,\rho^{(2)}(\Omega;\omega_1,\omega_2)
			+\frac{i}{\hbar}[H_0,\rho^{(2)}(\Omega;\omega_1,\omega_2)]
			\\
			&=
			-\frac{i}{\hbar}\Big(
			[H_{\mathrm E}(\omega_1),\rho^{(1)}(\omega_2)]
			+
			[H_{\mathrm E}(\omega_2),\rho^{(1)}(\omega_1)]
			\Big).   
		\end{split}
		\label{eq:rho2_freq_eom}
	\end{equation}
	We retain both ordered contributions in Eq.~\eqref{eq:rho2_freq_eom} so that the second-order kernel treats the two input frequencies on equal footing. The corresponding symmetry between the associated Cartesian field indices becomes explicit once the commutators are evaluated in the Bloch basis.
	
	To solve Eqs.~\eqref{eq:rho1_freq_eom} and \eqref{eq:rho2_freq_eom} in the Bloch basis, one must evaluate the position operator $x^a$, which is subtle in a periodic crystal. Following the work by Blount~\cite{BLOUNT1962,Aversa1995,Sipe2000}, we decompose it into intraband and interband parts,
	\begin{equation}
		x^a = R^a + r^a.
		\label{eq:Blount}
	\end{equation}
	In the Bloch-band representation, the interband part is purely off-diagonal,
	\begin{equation}
		r^a_{mn}=(1-\delta_{mn})\mathcal{A}^a_{mn},
		\label{eq:r_inter}
	\end{equation}
	while the intraband part is band diagonal,
	\begin{equation}
		R^a_{mn}=\delta_{mn}\left(i\partial_{a}+\mathcal{A}^a_{nn}\right),
		\label{eq:R_intra}
	\end{equation}
	where
	\begin{equation}
		\mathcal{A}^a_{mn}=i\langle u_{m\mathbf{k}}|\partial_{a}u_{n\mathbf{k}}\rangle
		\label{eq:Berry_connection}
	\end{equation}
	is the Berry-connection matrix and $\partial_a\equiv\partial_{k_a}$. 
	
	Because the intraband operator $R^a$ contains a momentum derivative, commutators involving $R^a$ are most naturally expressed through the generalized covariant derivative~\cite{Aversa1995,Sipe2000}. For any operator $S$, we define
	\begin{equation}
		S^{;a}_{nm}
		\equiv i[S,R^a]_{nm}
		=
		\partial_{a}S_{nm}
		-iS_{nm}\left(\mathcal{A}^a_{nn}-\mathcal{A}^a_{mm}\right).
		\label{eq:gen_deriv}
	\end{equation}
	In particular, for the interband position matrix $r^a_{mn}$, which is understood to refer to the off-diagonal sector $m\neq n$, the generalized derivative is
	\begin{equation}
		r^{a;b}_{nm}
		\equiv i[r^a,R^b]_{nm}
		=
		\partial_{b}r^a_{nm}
		-ir^a_{nm}\left(\mathcal{A}^b_{nn}-\mathcal{A}^b_{mm}\right).
		\label{eq:r_gen_deriv}
	\end{equation}
	Equations~\eqref{eq:r_inter} and \eqref{eq:r_gen_deriv} are written in a band-resolved Abelian
	notation and therefore assume that individual bands can be treated as
	nondegenerate. For a symmetry-enforced degeneracy, the degenerate multiplet
	should instead be treated as a single energy subspace, with the Berry
	connection and generalized derivative promoted to their matrix-valued
	non-Abelian counterparts.
	
	As detailed in Appendix~\ref{appendix:rho1_length}, the first-order density matrix in the Bloch-band basis contains both diagonal and off-diagonal terms,
	\begin{equation}
		\rho^{(1)}_{mn}(\omega_1)=\frac{-qE^b(\omega_1)}{\hbar}\big(\frac{i\partial_bf_n}{\omega_1}\delta_{mn}+\frac{r^b_{mn}f_{mn}}{\omega_{mn}-\omega_1}
		\big),
		\label{eq:rho1_main}
	\end{equation}
	where $f_{mn}\equiv f_m-f_n$ and $\omega_{mn}\equiv(\epsilon_m-\epsilon_n)/\hbar$. 
	
	\begin{figure}[htbp]
		\includegraphics[width=1.\columnwidth]{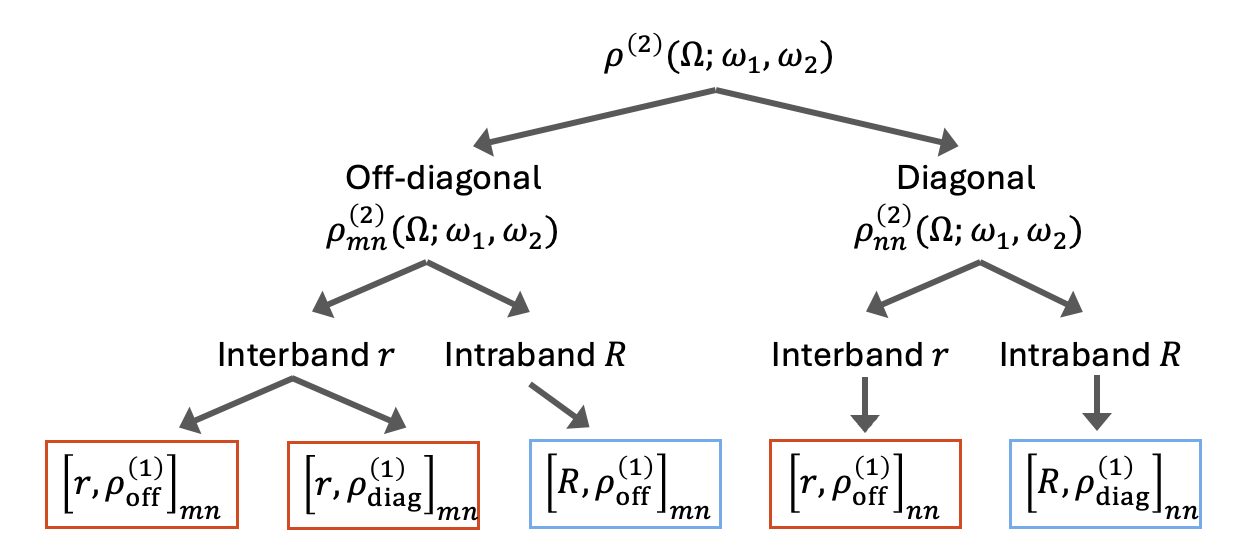}
		\caption{
			Tree representation of the second-order density matrix in the length gauge.
			The response first separates into off-diagonal coherences
			\(\rho^{(2)}_{mn}\) with \(m\neq n\) and the diagonal population response
			\(\rho^{(2)}_{nn}\). Each sector is generated by interband position matrix
			elements \(r\) and intraband Berry connection derivatives \(R\), acting on
			the first-order density matrix. Red boxes denote interband coupling
			contributions, while blue boxes denote intraband derivative contributions.
			This organization is natural in the length gauge because the current operator
			is field independent and all field dependence resides in the density matrix.
		}
		\label{fig1:rho2}
	\end{figure}
	The second-order response is governed by
	\begin{equation}
		\begin{split}
			i(\omega_{mn}-\Omega)\rho^{(2)}_{mn}(\Omega;\omega_1,\omega_2)
			&=\frac{iq}{\hbar}\Big(
			E^b(\omega_1)[x^b,\rho^{(1)c}(\omega_2)]_{mn}\\
			&+
			E^c(\omega_2)[x^c,\rho^{(1)b}(\omega_1)]_{mn}
			\Big),
		\end{split}
		\label{eq:rho2_band_start}
	\end{equation}
	with $\Omega=\omega_1+\omega_2$. For compactness, we use the shorthand $(b,\omega_1)\leftrightarrow(c,\omega_2)$ to denote the additional term obtained by simultaneously interchanging the Cartesian indices and input frequencies in the preceding expression.

	Because $\rho^{(1)}$ contains both diagonal and off-diagonal parts, and the position operator decomposes as $x^a=r^a+R^a$, the evaluation of $\rho^{(2)}$ naturally separates into the branches summarized in Fig.~\ref{fig1:rho2}. Details are given in Appendix~\ref{appendix:rho2_length}.
	
	For $m\neq n$, the second-order density matrix can be written as
	\begin{equation}
		\begin{split}
			\rho^{(2)}_{mn}(\Omega;\omega_1,\omega_2)
			&=
			-\frac{q^2E^b(\omega_1)E^c(\omega_2)}{\hbar^2(\omega_{mn}-\Omega)}
			\Big(
			\mathcal{K}^{bc}_{mn}(\omega_2)\\
			&+(b,\omega_1)\leftrightarrow(c,\omega_2)
			\Big),
			\label{eq:rho2_offdiag_main}
		\end{split}
	\end{equation}
	where $\mathcal{K}^{bc}_{mn}(\omega)$ collects the contributions generated by both the off-diagonal and diagonal parts of $\rho^{(1)}$:
	\begin{equation}
		\begin{split}
			&\mathcal{K}^{bc}_{mn}(\omega)\equiv i\,r^{c;b}_{mn}\frac{f_{mn}}{\omega_{mn}-\omega}
			+i\,r^c_{mn}\partial_b\left(\frac{f_{mn}}{\omega_{mn}-\omega}\right)\\
			&+\sum_{\ell\neq m,n}\Big(
			r^b_{m\ell}r^c_{\ell n}\frac{f_{\ell n}}{\omega_{\ell n}-\omega}
			-r^{c}_{m\ell}r^b_{\ell n}\frac{f_{m\ell}}{\omega_{m\ell}-\omega}
			\Big)\\
			&-\frac{i}{\omega}r^b_{mn}\partial_cf_{mn}.   
		\end{split}   
		\label{eq:Kappa}
	\end{equation}
	In the absence of the diagonal first-order contribution, the off-diagonal second-order density matrix reduces to the standard length gauge form of Sipe~\cite{Aversa1995,Sipe2000}. The additional terms proportional to $\partial_c f_{mn}$ arise from retaining the band-diagonal first-order density matrix~\cite{Jia2024} and encode the Fermi surface contribution relevant in the DC limit.

	The diagonal second-order density matrix is
	\begin{equation}
		\begin{split}
			&\rho^{(2)}_{nn}(\Omega;\omega_1,\omega_2)
			=
			\frac{q^2E^b(\omega_1)E^c(\omega_2)}{\hbar^2\Omega}
			\Bigg[
			-\frac{\partial_b\partial_c f_n}{\omega_2}\\
			&+\sum_{\ell\neq n}f_{\ell n}
			\left(
			\frac{r^b_{n\ell}r^c_{\ell n}}{\omega_{\ell n}-\omega_2}
			-\frac{r^c_{n\ell}r^b_{\ell n}}{\omega_{\ell n}+\omega_2}
			\right)
			+\,(b,\omega_1)\leftrightarrow(c,\omega_2)
			\Bigg].
		\end{split}
		\label{eq:rho2_diag_main}
	\end{equation}
	For clarity, we organize the second-order density matrix by separating its diagonal and off-diagonal sectors from the outset, which makes the origin of the different contributions explicit. Earlier length gauge formulations~\cite{Su2024,Jia2024} often group the same recursion according to covariant-derivative structures before this sector projection is made. The two organizations are equivalent; only the final conductivity tensor obtained after combining all contributions is physically meaningful.
	
	Having established the perturbative density matrix up to second-order in the electric field, we now turn to the corresponding current response. The electric current is obtained from the physical velocity operator,
	\begin{equation}
		J^a(t)=q\,\mathrm{Tr}\!\left[v^a\rho(t)\right],
		\qquad
		v^a=\frac{i}{\hbar}[H,x^a],
		\label{eq:J_def}
	\end{equation}
	where $H=H_0+H_{\mathrm E}$. In the length gauge, $H_{\mathrm E}(t)=-qE^b(t)x^b$, and since the position operators commute (see Appendix~\ref{appendix:position_algebra}), one has $[H_{\mathrm E},x^a]=0$. Therefore,
	\begin{equation}
		v^a=\frac{i}{\hbar}[H_0,x^a],
		\label{eq:v_H0_only}
	\end{equation}
	so that the electric field enters the current entirely through the perturbative density matrix.
	
	This feature distinguishes the present formulation from approaches that decompose the response into intraband and interband polarization channels, such as the Aversa-Sipe construction~\cite{Aversa1995,Sipe2000}. In that bookkeeping, additional terms appear in the corresponding channel-resolved effective currents, even though the total physical current is still generated by Eq.~\eqref{eq:v_H0_only}; see Appendix~\ref{append:intra} for detailed discussions. It should also be contrasted with the velocity gauge, where the minimal-coupling Hamiltonian is expanded in powers of the vector potential and the nonlinear current is correspondingly organized into separate contributions from the expanded current operator and the expanded density matrix.
	
	We now expand the total current order by order in the electric field,
	\begin{equation}
		J^a(t)=J^{a,(0)}+J^{a,(1)}(t)+J^{a,(2)}(t)+\cdots,
	\end{equation}
	with
	\begin{equation}
		J^{a,(n)}(t)=q\,\mathrm{Tr}\!\left[v^a\rho^{(n)}(t)\right].
		\label{eq:J_n_def}
	\end{equation}
	In frequency space, these define the corresponding response tensors. At linear order, the present formalism reproduces the standard conductivity tensor, including the clean-limit Drude contribution and the intrinsic anomalous Hall term in the dc limit. Since these results are well known, we summarize them in Appendix~\ref{appendix:linear_response} and focus here on the second-order response.
	\subsection{Second-order conductivity tensor}
	
	We now turn to the central quantity of interest, the second-order conductivity tensor. In frequency space, the second-order current is written as
	\begin{equation}
		\begin{split}
			J^{a,(2)}(\Omega)
			=&
			\int\frac{d\omega_1}{2\pi}\frac{d\omega_2}{2\pi}\,
			(2\pi)\delta(\Omega-\omega_1-\omega_2)\,\\
			&\chi^{abc}(\Omega;\omega_1,\omega_2)\,
			E^b(\omega_1)E^c(\omega_2).
		\end{split}
		\label{eq:J2_chi_def}
	\end{equation}
	Using
	\begin{equation}
		J^{a,(2)}(\Omega)=q\,\mathrm{Tr}\!\left[v^a\rho^{(2)}(\Omega)\right],
		\label{eq:J2_rho2}
	\end{equation}
	the conductivity tensor is obtained as
	\begin{equation}
		\chi^{abc}(\Omega;\omega_1,\omega_2)
		=
		\frac{q\,\mathrm{Tr}\!\left[v^a\rho^{(2),bc}(\Omega;\omega_1,\omega_2)\right]}
		{E^b(\omega_1)E^c(\omega_2)},
		\label{eq:chi_from_rho2}
	\end{equation}
	where $\rho^{(2),bc}$ denotes the component of the second-order density matrix generated by the fields $E^b(\omega_1)$ and $E^c(\omega_2)$, as given in Eqs.~\eqref{eq:rho2_offdiag_main} and \eqref{eq:rho2_diag_main}.
	
	Separating the trace into off-diagonal and diagonal sectors, we obtain
	\begin{equation}
		\begin{split}
			\chi^{abc}(\Omega;\omega_1,\omega_2)
			&=
			q\sum_{n,m\neq n} v^a_{nm}\,
			\frac{\rho^{(2)}_{mn}(\Omega;\omega_1,\omega_2)}{E^b(\omega_1)E^c(\omega_2)}\\
			&\qquad
			+q\sum_n v^a_{nn}\,
			\frac{\rho^{(2)}_{nn}(\Omega;\omega_1,\omega_2)}{E^b(\omega_1)E^c(\omega_2)}.
		\end{split}
		\label{eq:chi02_def}
	\end{equation}
	
	For compactness, we suppress the full frequency arguments on $\chi^{abc}_{\rm od}$ and $\chi^{abc}_{\rm diag}$ below. The off-diagonal contribution is
	\begin{equation}
		\chi^{abc}_{\rm od}
		=
		-\frac{q^3}{\hbar^2}\sum_{n,m\neq n}
		\frac{v^a_{nm}}{\omega_{mn}-\Omega}
		\Big[
		\mathcal{K}^{bc}_{mn}(\omega_2)+\mathcal{K}^{cb}_{mn}(\omega_1)
		\Big].
		\label{eq:chi_od}
	\end{equation}
	The diagonal contribution is
	\begin{equation}
		\begin{split}
			&\chi^{abc}_{\rm diag}=\frac{q^3}{\hbar^2}\sum_n \frac{v^a_{nn}}{\Omega} \Bigg[\sum_{m\neq n}f_{mn}\Bigg(
			\frac{r^b_{nm}r^c_{m n}}{\omega_{m n}-\omega_2}
			-\frac{r^c_{nm}r^b_{m n}}{\omega_{m n}+\omega_2}\\
			&+\frac{r^c_{nm}r^b_{m n}}{\omega_{m n}-\omega_1}
			-\frac{r^b_{nm}r^c_{m n}}{\omega_{m n}+\omega_1}
			\Bigg)-\left(\frac{\partial_b\partial_c f_n}{\omega_2}
			+\frac{\partial_c\partial_b f_n}{\omega_1}
			\right)\Bigg].
		\end{split}
		\label{eq:chi_diag}
	\end{equation}
	At this stage, both the off-diagonal and diagonal sectors contain terms that are naturally associated with Fermi surface response, as well as terms that appear to define a finite and gauge-invariant Fermi sea contribution. The physically relevant question is whether such Fermi sea terms survive in the strict adiabatic dc limit after all sectors are combined.
	
	\subsection{Static nonlinear conductivity}
	
	\subsubsection{Adiabatic dc limit and cancellation of Fermi sea terms}
	To obtain the strict dc response, we apply the adiabatic prescription consistent with the Fourier convention $e^{-i\omega t}$~\cite{Das2023}, namely
	\begin{equation}
		\omega_1\to i\eta,\quad
		\omega_2\to i\eta,\quad
		\Omega\to 2i\eta,
		\quad
		\eta\to0^+.
		\label{eq:dc_prescription_second}
	\end{equation}
	A detailed derivation is given in Appendix~\ref{appendix:2nd_response}, but the cancellation mechanism can be summarized briefly. In the strict adiabatic dc limit, both the diagonal and off-diagonal sectors generate terms that appear to define finite Fermi sea contributions. After combining the generalized-derivative structures from the two sectors, one obtains
	\begin{equation}
		\chi^{abc}_{\rm sea,gd}
		=
		\sum_{m,n}F_{mn}\Big[
		r^c_{mn}\big(r^{a;b}_{nm}-r^{b;a}_{nm}\big)
		+r^b_{mn}\big(r^{a;c}_{nm}-r^{c;a}_{nm}\big)
		\Big],
		\label{eq:sea_combined_commutator_form}
	\end{equation}
	while the remaining off-diagonal contribution is a triple-\(r\) term involving an intermediate band index,
	\begin{equation}
		\begin{split}
			\chi^{abc}_{\mathrm{od},r}
			&=
			\sum_{\ell,m,n} i\Big(
			r^a_{nm}r^b_{m\ell}r^c_{\ell n}F_{\ell n}
			-r^a_{nm}r^{c}_{m\ell}r^b_{\ell n}F_{m\ell}
			+b\leftrightarrow c
			\Big)
			\\
			&=
			\sum_{m,n} F_{mn}\Big(
			r^c_{mn}\,i[r^a,r^b]_{nm}
			+r^b_{mn}\,i[r^a,r^c]_{nm}
			\Big)
			\notag\\
			&=
			-\sum_{m,n}F_{mn}\Big[
			r^c_{mn}\big(r^{a;b}_{nm}-r^{b;a}_{nm}\big)
			+r^b_{mn}\big(r^{a;c}_{nm}-r^{c;a}_{nm}\big)
			\Big].
		\end{split}
		\label{eq:triple_r_rewrite}
	\end{equation}
	Here we suppress the common prefactor \(q^3/\hbar^2\) and define \(F_{mn}\equiv f_{mn}/\omega_{mn}\) for brevity. Here \(F_{mn}\) is defined only for \(m\neq n\), and all sums involving \(r^a_{mn}\) are correspondingly restricted to interband indices. In particular, \(\sum_{m\neq n}\) denotes an interband double sum, while the intermediate-band sum \(\sum_{\ell\neq m,n}\) excludes \(\ell=m,n\).
	Thus, in the strict adiabatic dc limit, the triple-\(r\) term is exactly the negative of the generalized-derivative contribution, and all apparent Fermi sea terms cancel. 
	
	This result also clarifies the status of the apparent Fermi sea term identified in previous velocity gauge analyses~\cite{Kaplan2023}: no such term survives in the strict adiabatic dc limit. As we show later, the velocity gauge formulation exhibits the same cancellation, as required by the gauge invariance of the physical response.
	
	\subsubsection{Surviving Fermi surface contribution}
	
	After the cancellation of all apparent Fermi sea terms, the strict adiabatic dc response is purely Fermi surface in origin. The resulting static nonlinear conductivity can be written as
	\begin{equation}
		\chi^{abc}_{\rm dc}
		=
		\frac{q^3}{\hbar}
		\sum_n\left(
		G^{ab}_n\partial_c f_n
		+
		G^{ca}_n\partial_b f_n
		-\frac{1}{2}G^{bc}_n\partial_a f_n
		\right),
		\label{eq:chi_dc_real_explicit}
	\end{equation}
	where
	\begin{equation}
		G^{ab}_n
		\equiv
		\sum_{m\neq n}\frac{2g^{ab}_{nm}}{\epsilon_{nm}}
		=
		2\,\mathrm{Re}\sum_{m\neq n}\frac{r^a_{nm}r^b_{mn}}{\epsilon_{nm}}
		\label{eq:BCP_def_main}
	\end{equation}
	is the Berry connection polarizability tensor~\cite{Gao2014,Xiao2021,Yang2021} and 
	\begin{equation}
		g^{ab}_{nm}\equiv \frac{1}{2}\left(r^a_{nm}r^b_{mn}+r^b_{nm}r^a_{mn}\right)
		\label{eq:g_nm_main}
	\end{equation}
	is the symmetrized interband quantum metric tensor. 
	For an exactly degenerate multiplet, Eq.~\eqref{eq:chi_dc_real_explicit} is understood in the
	corresponding subspace form: only transitions between distinct energy
	subspaces enter the interband denominator, and the physical conductivity
	involves the trace of the resulting matrix-valued BCP over the degenerate
	subspace~\cite{Yang2021}.
	
	Equation~\eqref{eq:chi_dc_real_explicit} summarizes the main implication of the present length gauge analysis. Although the diagonal and off-diagonal sectors each generate apparently finite Fermi sea structures at intermediate stages, these cancel exactly in the strict adiabatic dc limit. The resulting static nonlinear conductivity is therefore purely Fermi surface in origin, consistent with the Berry-connection-polarizability form and incompatible with an intrinsic dc response in an insulator.

	Equation~\eqref{eq:chi_dc_real_explicit} can be decomposed into dissipative and reactive parts according to whether it contributes to the dissipation rate
	\begin{equation}
		\dot Q = J^a_{\rm dc}E^a = \chi^{abc}_{\rm dc}E^aE^bE^c.
	\end{equation}
	Since \(E^aE^bE^c\) is fully symmetric under permutations of \(a,b,c\), only the fully symmetrized part of \(\chi^{abc}_{\rm dc}\) contributes to dissipation. We therefore write~\cite{Souza2022}
	\begin{equation}
		\chi^{abc}_{\rm dc}=\chi^{abc}_{\rm diss}+\chi^{abc}_{\rm rea},
		\label{eq:chi_dc_split_main}
	\end{equation}
	where
	\begin{equation}
		\chi^{abc}_{\rm diss}
		\equiv
		\frac{1}{6}\Big(
		\chi^{abc}_{\rm dc}
		+\chi^{acb}_{\rm dc}
		+\chi^{bac}_{\rm dc}
		+\chi^{bca}_{\rm dc}
		+\chi^{cab}_{\rm dc}
		+\chi^{cba}_{\rm dc}
		\Big)
		\label{eq:chi_diss_main}
	\end{equation}
	is the dissipative part and \(\chi^{abc}_{\rm rea}\equiv \chi^{abc}_{\rm dc}-\chi^{abc}_{\rm diss}\) is the reactive part.
	Using the symmetry \(G_n^{ab}=G_n^{ba}\), one obtains
	\begin{equation}
		\begin{aligned}
			\chi^{abc}_{\rm diss}
			&=
			\frac{q^3}{2\hbar}\sum_n\left(
			G^{ab}_n\partial_c f_n
			+
			G^{ca}_n\partial_b f_n
			+
			G^{bc}_n\partial_a f_n
			\right),\\
			\chi^{abc}_{\rm rea}
			&=
			\frac{q^3}{2\hbar}\sum_n\left(
			G^{ab}_n\partial_c f_n
			+
			G^{ca}_n\partial_b f_n
			-
			2G^{bc}_n\partial_a f_n
			\right).
		\end{aligned}
		\label{eq:chi_diss_rea_main}
	\end{equation}
	
	The decomposition in Eq.~\eqref{eq:chi_diss_rea_main} also clarifies the physical content of the static response. The reactive component \(\chi^{abc}_{\rm rea}\) is the proper intrinsic second-order anomalous Hall response~\cite{Gao2014}, whereas the dissipative component \(\chi^{abc}_{\rm diss}\) gives an intrinsic nonlinear Ohmic current~\cite{Das2023,Kaplan2024,Su2024,Jia2024}. While different names have been used in the literature for the latter, both effects share the same symmetry requirements: they vanish if either time-reversal or inversion is preserved, but can remain finite when the combined $\mathcal{PT}$ symmetry is present. Consequently, these intrinsic nonlinear dc responses provide a powerful probe of Néel order in $\mathcal{PT}$-symmetric antiferromagnets~\cite{Xiao2021,Yang2021}.

	\section{Quantum Kinematics in the Velocity Gauge}\label{sec:velocity_gauge}
	
\begin{figure}[htbp]
		\includegraphics[width=1.\columnwidth]{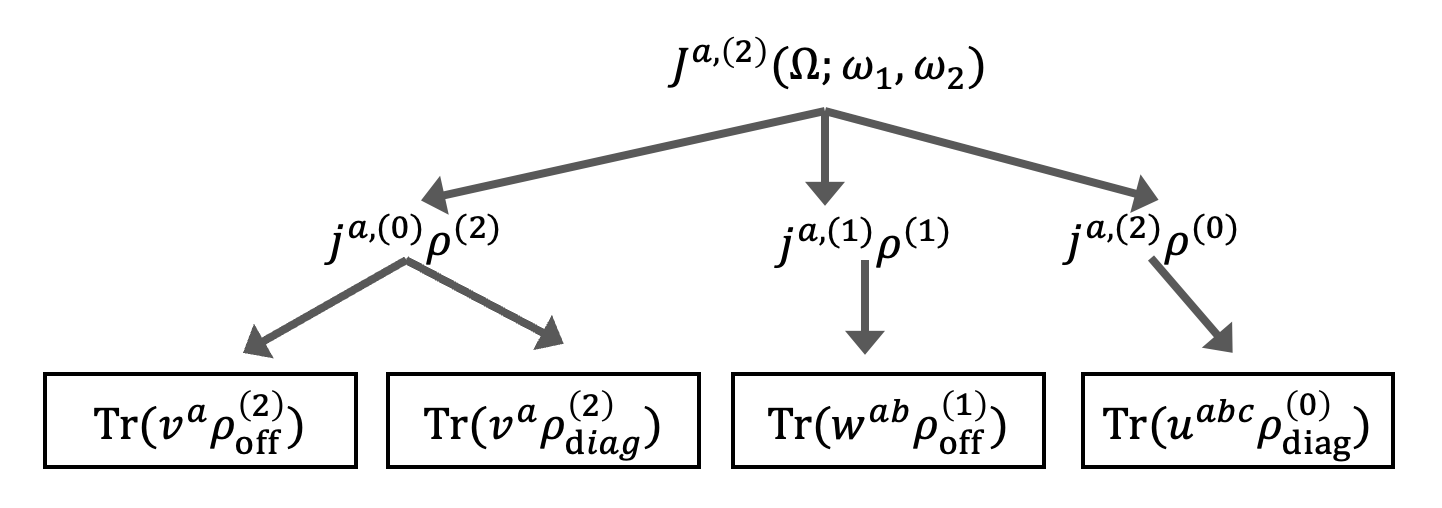}
		\caption{
			Tree representation of the second-order current in the velocity gauge. Unlike
			the length gauge, the current operator itself contains field-dependent terms,
			so \(J^{a,(2)}\) receives contributions from
			\(j^{a,(0)}\rho^{(2)}\), \(j^{a,(1)}\rho^{(1)}\), and
			\(j^{a,(2)}\rho^{(0)}\). The first branch further separates into off-diagonal
			and diagonal parts of \(\rho^{(2)}\). Constants and external vector potentials
			multiplying \(v^a\), \(w^{ab}\), and \(u^{abc}\) are suppressed. The individual
			branches are bookkeeping terms only; the length gauge response is recovered
			after all branches are combined.
		}
		\label{fig2:Jv}
	\end{figure}

	A gauge invariant formulation is essential for obtaining a unique nonlinear
	conductivity from quantum perturbation theory. The same electric field can be
	introduced either through the scalar potential coupling
	\(-q\mathbf{E}\cdot\mathbf{r}\), as in the length gauge, or through minimal
	coupling \(\mathbf{k}\rightarrow \mathbf{k}-q\mathbf{A}(t)/\hbar\), as in the
	velocity gauge~\cite{Luttinger1955,Passos2018,Yan2023,Kaplan2023,Kaplan2024}. Although these two representations are physically equivalent,
	they organize the perturbation expansion in very different ways. In the
	length gauge, the current operator is field independent and the response is
	generated by the field-induced density matrix, as summarized in
	Fig.~\ref{fig1:rho2}. In the velocity gauge, by
	contrast, both the Hamiltonian and the current operator acquire explicit
	field-dependent terms, leading to
	the decomposition shown in Fig.~\ref{fig2:Jv}. Demonstrating the equivalence of the two formulations
	therefore provides a stringent check on the density-matrix approach and on
	the gauge invariance of the resulting nonlinear conductivity.
	We now derive the same nonlinear response in the velocity gauge. This
	provides an independent check on the length gauge result and clarifies the
	role of the field-dependent current operator. 
	In the velocity gauge, the external electric field is introduced by the
	minimal substitution
	\begin{equation}
		H(\mathbf{k},t)=H_0\!\left(\mathbf{k}-\frac{q}{\hbar}\mathbf{A}(t)\right),
	\end{equation}
	where \(q\) is the carrier charge and
	\(\mathbf{E}(\omega)=i\omega \mathbf{A}(\omega)\). Expanding in powers of
	\(\mathbf{A}\), we write
	\begin{equation}
		\begin{split}
			H(t)
			&=H_0-qA^b(t)v^b
			+\frac{q^2}{2}A^b(t)A^c(t)w^{bc}  \\
			&\quad
			-\frac{q^3}{3!}A^b(t)A^c(t)A^d(t)u^{bcd}
			+\cdots ,
		\end{split}
	\end{equation}
	with
	\begin{equation}
		v^b=\frac{1}{\hbar}\partial_b H_0,\quad
		w^{bc}=\frac{1}{\hbar^2}\partial_b\partial_c H_0,\quad
		u^{bcd}=\frac{1}{\hbar^3}\partial_b\partial_c\partial_d H_0 .
	\end{equation}
	The corresponding current operator is
	\begin{equation}
		j^a(t)=-\frac{\partial H}{\partial A^a(t)} .
	\end{equation}
	It therefore contains explicit field-dependent pieces,
	\begin{equation}
		j^a(t)
		=
		qv^a
		-q^2A^b(t)w^{ab}
		+\frac{q^3}{2}A^b(t)A^c(t)u^{abc}
		+\cdots .
	\end{equation}
	
	The full finite frequency velocity gauge expression is algebraically lengthy
	because the same length gauge structures are distributed among
	\(j^{(0)}\rho^{(2)}\), \(j^{(1)}\rho^{(1)}\), and \(j^{(2)}\rho^{(0)}\) (Fig.~\ref{fig2:Jv}). We
	therefore focus here on the static adiabatic limit and leave the intermediate
	steps to Appendix~\ref{app:conductivity_velocity_gauge}. Along the trajectory
	\(\omega_1=\omega_2=i\eta\), \(\Omega=2i\eta\), we expand the full
	velocity gauge response and retain the finite \(\eta^0\) coefficient. Up to the common prefactor \(q^3/\hbar^2\), this gives
	
	\begin{widetext}
		\begin{equation}
			\begin{split}
				\chi^{abc}_{\rm VG}&=
				\sum_{\substack{n,m \neq n \\ \ell \neq m,n}}
				\left[
				f_{\ell n}v^a_{nm}v^b_{m\ell}v^c_{\ell n}
				\left(
				\frac{1}{\omega_{mn}\omega_{\ell n}^3}
				+\frac{2}{\omega_{mn}^2\omega_{\ell n}^2}
				+\frac{4}{\omega_{mn}^3\omega_{\ell n}}
				\right)-
				f_{m\ell}v^a_{nm}v^c_{m\ell}v^b_{\ell n}
				\left(\frac{1}{\omega_{mn}\omega_{m\ell}^3}+\frac{2}{\omega_{mn}^2\omega_{m\ell}^2}+\frac{4}{\omega_{mn}^3\omega_{m\ell}}\right)\right]\\
				&+\sum_{n,m\neq n}f_{mn}\frac{7v^a_{nm}\Delta^b_{mn}v^c_{mn}}
				{\omega_{mn}^4}
				-2f_{mn}\frac{v^a_{nm}\hbar w^{bc}_{mn}}{\omega_{mn}^3}
				+f_{mn}\frac{\Delta^a_{mn}v^b_{nm}v^c_{mn}}
				{2\omega_{mn}^4}
				-f_{mn}\frac{\hbar w^{ba}_{nm}v^c_{mn}}
				{\omega^3_{mn}}
				+
				(b\leftrightarrow c).
			\end{split}
			\label{eq:chi_static_vg}
		\end{equation}
	\end{widetext}
	
	The first line contains the genuine three-band terms from the off-diagonal second-order density matrix. The second line contains the two-band terms
	generated by endpoint contributions and by the field-dependent current
	operators. Although these pieces do not resemble the length gauge result term by term, using the matrix element identity
	for \(w^{ab}\), the Jacobi identity for the triple-\(r\) sector, and integration by parts over the Brillouin zone, Eq.~\eqref{eq:chi_static_vg} reduces  exactly to the length gauge static result in Eq.~\eqref{eq:fermi surface length}.
	This agreement provides a nontrivial gauge equivalence check: the apparent
	Fermi sea pieces in the velocity gauge cancel only after the
	field-dependent-current branches \(j^{(1)}\rho^{(1)}\) and
	\(j^{(2)}\rho^{(0)}\) are included.
	
	\subsection{Two-band model demonstration}
	
	As a direct numerical check of the gauge equivalence, we consider a tilted
	two-band lattice massive Dirac model,
	\begin{equation}
		\begin{split}
			H(\mathbf{k})
			=
			&t\sin k_x\,\sigma_0
			+v\sin k_x\,\sigma_x
			+v\sin k_y\,\sigma_y\\
			&+\left[M+B(2-\cos k_x-\cos k_y)\right]\sigma_z . 
		\end{split}
		\label{eq:two_band_model}
	\end{equation}
	We evaluate the nonlinear conductivity $\chi^{xyy}$ independently using the
	length gauge expression in Eq.~\eqref{eq:fermi surface length} and the
	velocity gauge expression in Eq.~\eqref{eq:chi_static_vg}. Since the model
	contains only two bands, the genuine three-band contribution in the velocity
	gauge vanishes identically, providing a particularly simple test of the
	remaining two-band terms.
	
	\begin{figure}[htbp]
		\centering\
		\includegraphics[width=1.\columnwidth]{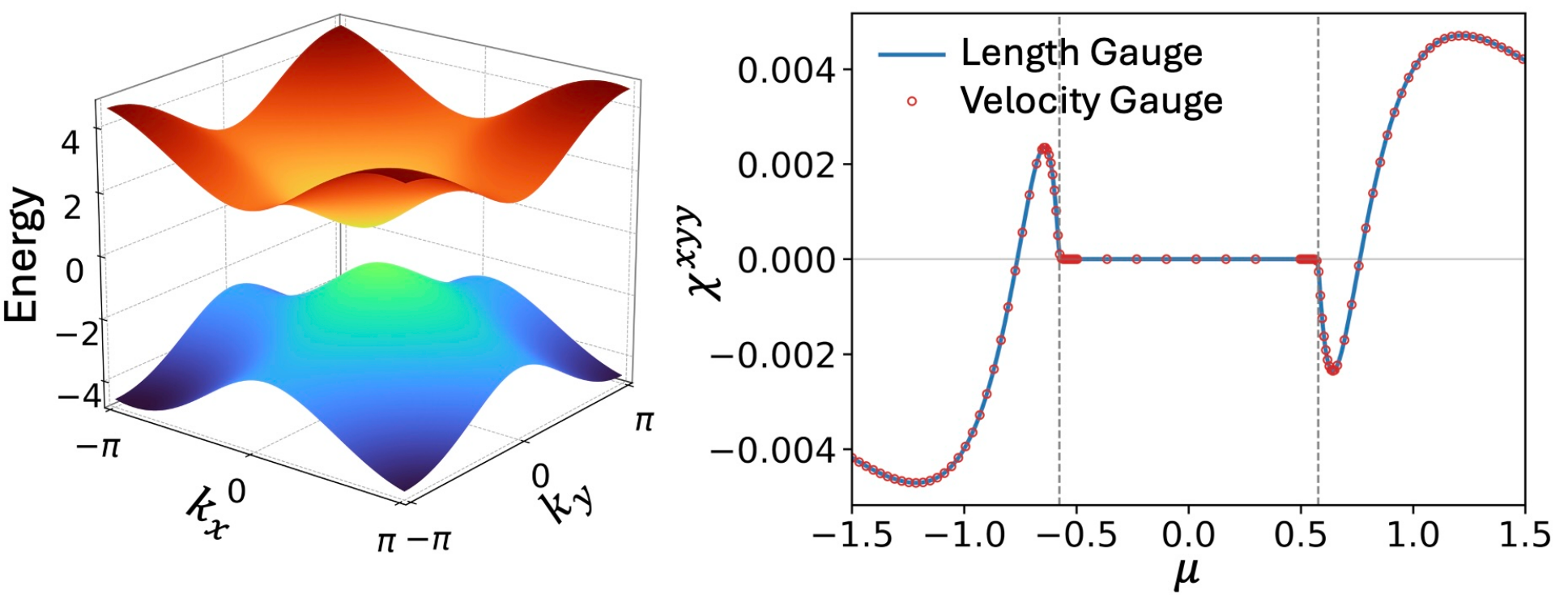}
		\caption{
			Numerical demonstration of the length-velocity gauge equivalence for the
			two-band lattice model in Eq.~\eqref{eq:two_band_model}. Left shows band dispersion over the full Brillouin zone.
			Right shows nonlinear conductivity $\chi^{xyy}$ as a function of chemical potential
			$\mu$, evaluated independently in the length gauge (solid line) and velocity
			gauge (open circles). The vertical dashed lines indicate the valence-band
			maximum and conduction-band minimum. We use $t=0.35$, $v=1$, $M=0.60$, $B=1$, $a=\hbar=q=1$, and
			$k_BT=0.003$.
			The two calculations coincide throughout the metallic regimes and both
			vanish when the chemical potential lies inside the gap.
		}
		\label{fig:two_band_model}
	\end{figure}
	
	As shown in Fig.~\ref{fig:two_band_model}, the independently evaluated
	length and velocity gauge conductivities agree within numerical precision
	over the full range of chemical potential considered. In particular, both
	formulations give a vanishing response in the insulating gap and the same
	Fermi-surface response once the chemical potential enters either band. This
	provides a direct model-level verification of the analytical gauge
	equivalence established above.
	\section{Discussion}\label{sec:discussion}
	
	Our central result is that the intrinsic second-order static response is
	gauge independent when the adiabatic limit is taken consistently. In the length gauge, the response is organized by the diagonal and off-diagonal components of the density matrix, together with the distinction between interband position-matrix elements \(r\) and intraband covariant derivatives \(R\). In the velocity gauge, the same response is
	distributed among \(j^{(0)}\rho^{(2)}\), \(j^{(1)}\rho^{(1)}\), and
	\(j^{(2)}\rho^{(0)}\). These individual pieces are not separately physical:
	they contain apparent Fermi sea and Fermi surface contributions whose form
	depends on the gauge and on the representation of the matrix elements
	\(w^{ab}\) and \(u^{abc}\). After all current-operator branches are included,
	the apparent Fermi sea terms cancel and the velocity gauge reduces exactly to
	the length gauge result.
	
	This cancellation has an immediate physical implication. The clean adiabatic
	dc nonlinear Hall response derived here is a Fermi surface quantum geometric
	term. Therefore it vanishes in a fully gapped insulator when the chemical
	potential lies in the gap. In this sense, there is no residual intrinsic
	Fermi sea dc nonlinear Hall current in the gauge-consistent static limit. This
	point is useful for interpreting quantum geometric nonlinear responses:
	relaxation-time independence alone does not imply an independent insulating dc
	transport contribution. The character of the response depends on the limiting
	procedure and on whether the term belongs to the clean retarded response or to
	a scattering-controlled transport current.
	
	\begin{table*}[t]
		\caption{
			Comparison of representative approaches to intrinsic second-order dc
			nonlinear conductivity. LG and VG denote the length and velocity gauges,
			respectively; WPE denotes the wave-packet-energy contribution; RTA denotes
			the relaxation-time approximation; and NESS denotes a nonequilibrium steady
			state. ``Adiabatic dc'' refers to the prescription
			$\omega_1=\omega_2=i\eta$, $\Omega=2i\eta$, followed by $\eta\to0^+$.
			The ``Fermi sea'' column indicates whether a nonzero intrinsic dc current
			survives when the chemical potential lies in a full gap. 
			The asterisk denotes agreement after the factor correction discussed in the text.
		}
		
		\label{tab:literature_comparison}
		\centering
		\footnotesize
		\setlength{\tabcolsep}{5pt}
		\renewcommand{\arraystretch}{1.15}
		\begin{tabular}{lllll}
			\hline\hline
			Reference
			& Formalism
			& DC prescription / relaxation
			& Fermi sea
			& Relation to present result
			\\
			\hline
			
			Refs.~\cite{Gao2014,Xiao2021,Yang2021}
			& Semiclassical wave packet
			& Static dc; clean
			& Absent
			& Same reactive; dissipative not addressed
			\\
			Ref.~\cite{Das2023}
			& LG density matrix
			& Adiabatic dc; effective $\tau/2$; clean
			& Absent
			& Same
			\\
			Ref.~\cite{Su2024}
			& LG density matrix
			& Adiabatic dc; clean
			& Absent
			& Same
			\\
			Ref.~\cite{Jia2024}
			& LG density matrix / WPE
			& Adiabatic dc; clean
			& Absent
			& LG: same$^*$; WPE: dissipative part differs
			\\
			Ref.~\cite{Kaplan2023}
			& VG diagrammatic
			& Adiabatic dc; clean
			& Present
			& Different
			\\
			
			Ref.~\cite{Balents2026}
			& Moyal quantum kinetics
			& Low-$\omega$ dc limit; RTA, $\tau^{-1}\to0$
			& Absent
			& Different
			\\
			Ref.~\cite{Chang2026}
			& Open-system NESS
			& $\omega\to0$; microscopic bath, $\Gamma^0$
			& Absent
			& Different
			\\
			Ref.~\cite{ulrich2025quantumgeometricoriginintrinsic}
			& Keldysh Green function
			& Static dc; constant $i\Gamma$, intrinsic $\tau^0$
			& Absent
			& Different
			\\
			
			Present work
			& LG + VG density matrix
			& Adiabatic dc; clean
			& Absent
			& ---
			\\
			\hline\hline
			
		\end{tabular}
		
	\end{table*}

	We compare our result with several recent quantum approaches to nonlinear
	quantum geometric transport~\cite{Das2023,Su2024,Kaplan2023,Kaplan2024,Jia2024,Liu2026Superadiabatic,Balents2026,ulrich2025quantumgeometricoriginintrinsic, Chang2026}.
	Table~\ref{tab:literature_comparison} summarizes the main differences in
	formalism, dc prescription, relaxation treatment, and the presence or absence
	of an intrinsic current in a full gap.
	Our result agrees with the length gauge density-matrix calculations of
	Refs.~\cite{Das2023,Su2024}. The intermediate derivations in these works are
	organized somewhat differently: Ref.~\cite{Das2023} introduces an effective
	relaxation time \(\tau/2\) in the dc expansion, while Ref.~\cite{Su2024} uses
	symmetry arguments to simplify the final expression. In contrast, our
	derivation keeps the full adiabatic expansion explicitly and shows that all
	apparent Fermi sea terms in the length gauge cancel after the off-diagonal and
	diagonal sectors are combined. The only surviving clean static nonlinear Hall
	contribution is the Fermi surface quantum metric term.
	
	The length gauge derivation of Ref.~\cite{Jia2024} is particularly close to
	ours. We find that their result becomes identical to ours, and to
	Refs.~\cite{Das2023,Su2024}, once a small numerical factor in the expansion of
	their ``ei'' contribution is corrected. Specifically, when the additional factor of two retained in our
	Eqs.~\eqref{eq:prefactor_expand_dc} and \eqref{eq:od_sur_raw} is included in the expansion associated with their Eqs.~(69) and (71), their result agrees with ours.

	Our velocity gauge derivation follows the same general logic as recent
	velocity gauge and diagrammatic approaches
	~\cite{Kaplan2023,Kaplan2024,Moore2019,Yan2020}, but makes the connection to
	the length gauge static limit explicit. In particular, Ref.~\cite{Kaplan2023}
	uses a different representation of the symmetric velocity gauge vertex
	\(w^{ab}=w^{ba}\). This choice redistributes terms between the two-band
	generalized-derivative sector and the triple-\(r\) sector, so the intermediate
	expressions can look different from ours. However, these representations are
	equivalent once the matrix-element sum rules, Jacobi identities, and
	Brillouin-zone integration by parts are applied, as shown in
	Appendix~\ref{app: symmetric wab}. Thus the properly reduced velocity gauge
	result agrees completely with the length gauge result: the apparent Fermi sea
	terms cancel in the clean static limit, and the surviving contribution is the
	same Fermi surface quantum-metric-dipole term.
	
	The Moyal-product quantum kinetic formulation of Ref.~\cite{Balents2026}
	also contains a Fermi surface contribution governed by the band-normalized
	quantum metric. Their Eq.~(3.35), however, is organized differently from
	our Eq.~\eqref{eq:chi_dc_real_explicit}, and the symmetrization over the two input fields is not displayed in the same form as in our density matrix formulation. As written, their quantum metric contribution therefore does not directly coincide with the coefficient structure of Eq.~\eqref{eq:chi_dc_real_explicit} although the two results clearly involve the same underlying quantum geometric structures. Their insulating response
	vanishes in the collisionless limit $\tau^{-1}\rightarrow 0$, which is the
	limit most directly comparable with the clean adiabatic response considered
	here.
	
	The comparison with semiclassical wave-packet theory requires more care. The
	reactive quantum geometric part of the response [Eq.~\eqref{eq:chi_diss_rea_main}] 
	is consistent with the semiclassical Berry-connection-polarizability picture of
	Refs.~\cite{Gao2014,Xiao2021,Yang2021}.  
	The dissipative/Ohmic sector,
	however, appears more sensitive to how the dc limit and relaxation are
	implemented. For example, the wave-packet-energy contribution proposed in
	Ref.~\cite{Jia2024} has, in our notation, the schematic tensor structure
	\begin{equation}
		\chi^{abc}_{\rm WPE}
		=\frac{1}{2}\left(
		G^{ac}\partial_b f+G^{ab}\partial_c f-
		G^{bc}\partial_a f\right),
	\end{equation}
	where \(G^{ab}\partial_c f\) denotes the band-normalized quantum metric
	structure with the Fermi surface derivative along \(c\). Applying the same
	cyclic decomposition used in Eq.~\eqref{eq:chi_diss_main}, this expression
	gives a dissipative/Ohmic part
	\begin{equation}
		\chi^{abc}_{\rm WPE,diss}
		=
		\frac{1}{6}
		\left(
		G^{ac}\partial_b f
		+
		G^{ab}\partial_c f
		+
		G^{bc}\partial_a f
		\right),
	\end{equation}
	and a complementary reactive part
	\begin{equation}
		\chi^{abc}_{\rm WPE,rea}
		=
		\frac{1}{3}
		\left(
		G^{ac}\partial_b f
		+
		G^{ab}\partial_c f
		-
		2G^{bc}\partial_a f
		\right).
	\end{equation}
	This partition differs from the coefficient structure obtained in the
	gauge-consistent density-matrix calculation and also does not reduce to the
	original semiclassical Berry-connection-polarizability form of
	Refs.~\cite{Gao2014,Xiao2021,Yang2021}. Thus the issue is not simply a discrepancy between
	semiclassical and quantum approaches. Rather, it raises a more basic question:
	how should the field-induced wave-packet energy correction be incorporated
	self-consistently with the relaxation process and the dc limiting procedure?
	We therefore view the dissipative/Ohmic sector as requiring further
	investigation, ideally within a kinetic framework that treats the wave-packet
	energy, collision integral, impurity scattering, and possible vertex
	corrections on the same footing~\cite{Du2021,Konig2021}.
	
	This question is closely related to recent Green-function, projector-based,
	and open-system formulations
	~\cite{ulrich2025quantumgeometricoriginintrinsic,Chang2026}.
	These approaches provide a natural language for treating quantum geometry
	together with relaxation, disorder, and nonequilibrium steady-state physics.
	In particular, Ref.~\cite{Chang2026} argues that the
	\(\Gamma^0\) nonlinear dc response is not a universal property of the Bloch
	Hamiltonian alone, but can depend on the dissipation mechanism that establishes
	the nonequilibrium steady state. This is consistent with our view that an
	\(\eta^0\) or \(\Gamma^0\) term should not automatically be interpreted as a
	disorder-free intrinsic dc transport coefficient.
	
	The distinction can be stated in terms of the density matrix being expanded.
	In the closed-system adiabatic response derived here, the only occupation
	input is the equilibrium distribution
	$\rho^{(0)}_{nn}=f_0(\epsilon_n)$. After combining the response contributions and integrating by parts, the finite quantum geometric response can be expressed entirely in terms of Fermi surface derivatives such as $\partial_a f_n$. The resulting band-normalized quantum metric term is thus obtained from the response built on the \textit{equilibrium} distribution. By contrast, in an open-system formulation,
	derivatives of $f_0$ can enter directly through the bath-determined perturbative steady-state density matrix. Such terms reflect the field-induced
	deformation of the steady-state occupation and therefore depend on the chosen
	dissipation mechanism.
	
	This distinction explains why the open-system result contains structures that
	partially overlap with ours, but does not reduce to our clean adiabatic
	Fermi-surface term. In the low-frequency SHG channel of
	Ref.~\cite{Chang2026}, the $T_0$-type contribution contains interband
	band-normalized quantum metric structures closely related to those appearing
	in our result. However, the full $\Gamma^0$ response is not simply this term
	written in another form. In their decomposition, the $T_0$ and $T_1$ terms
	together change the relative coefficients of the interband quantum metric
	structures, the $T_2$ term produces an intraband quantum-metric-derivative
	contribution, and the $T_3$ term gives a kinetic contribution associated
	with the bath-shaped nonequilibrium distribution. These additional terms
	should therefore be viewed as mechanism-dependent steady-state contributions,
	rather than as residual Fermi sea terms of the closed-system adiabatic
	response. A complete reconciliation of the dissipative nonlinear response
	therefore remains an important open problem.
	
	Although relaxation and collision effects are not included explicitly in the
	present calculation, their qualitative effects can be anticipated. In a
	simple relaxation-time or dephasing picture, finite decay rates regularize the nonlinear Drude-like poles and generate additional relaxation-dependent
	intraband contributions, while microscopic disorder can also produce
	extrinsic terms such as skew scattering and side jump~\cite{Du2021}. 
	
	Recovery of the clean quantum geometric contribution derived here requires a limiting procedure consistent with our adiabatic prescription. Specifically, in our clean calculation, both input fields are switched on with the same retarded adiabatic factor, giving $\omega_1=\omega_2=i\eta$. Frequency conservation then fixes the output frequency as $\Omega=\omega_1+\omega_2=2i\eta$. We expand the full response along this trajectory and retain the finite $\eta^0$ contribution as $\eta\to0^+$. This correlated continuation of the input and output frequencies is essential to the Fermi sea cancellation and the length–velocity gauge equivalence established here.
	
	More generally, the length and velocity gauges should remain physically equivalent in the presence of relaxation or disorder, provided the collision kernel or disorder self-energy, density matrix, and current operator are treated consistently under the gauge transformation. Phenomenological relaxation terms introduced independently in the two gauges need not preserve this covariance and can therefore generate apparent gauge-dependent differences. Related issues concerning relaxation and gauge equivalence in nonlinear optical response were discussed by Passos et al.~\cite{Passos2018}. Separately, different phenomenological relaxation prescriptions can yield different static nonlinear responses, as emphasized in Ref.~\cite{Yan2023}. Thus, a comparison of dc responses requires both a consistent limiting procedure and a gauge-consistent treatment of relaxation.

	To conclude, we have shown that the quantum density-matrix approach gives the
	same intrinsic second-order nonlinear conductivity in the length and velocity
	gauges when the adiabatic dc limit is taken consistently. The resulting clean
	static response is a Fermi surface contribution governed by the
	band-normalized quantum metric. Consequently, it vanishes in a fully gapped
	insulator when the chemical potential lies in the gap, confirming that there
	is no residual Fermi sea dc nonlinear Hall current in the gauge-consistent
	adiabatic limit.
	
	This result also clarifies the connection between nonlinear transport and quantum geometry.
	At linear order, the clean intrinsic Hall response is
	purely reactive and is governed by the Berry curvature. At second-order, the
	conductivity contains both reactive and dissipative components, and the
	relevant geometric structure is the band-normalized quantum metric. The
	reactive part obtained here agrees with the seminal semiclassical
	wave-packet result based on Berry-connection polarizability. By contrast, the
	dissipative part is more sensitive to how the dc limit, relaxation, and
	scattering are implemented. The remaining discrepancies among quantum
	density-matrix, semiclassical, and Green-function formulations suggest that a
	more complete treatment of impurity scattering and vertex corrections is
	needed to fully reconcile the Ohmic nonlinear response.
	Such a theory would place the reactive and dissipative quantum geometric
	responses on the same footing and provide a firmer basis for interpreting
	nonlinear Hall measurements in magnetic quantum materials.
	
	\section{Acknowledgments} 
	This work is supported by the National Science Foundation under Grant No.~OIA-2229498 and by UAB start-up funds. We thank Farzad Mahfouzi, Paul Haney, and Mark Stiles for fruitful discussions.
	
	\appendix
	
	\section{Density matrices in the length gauge} \label{appendix:rho_length}
	
	\subsection{First-order density matrix}\label{appendix:rho1_length}
	
	We work in the basis of cell-periodic Bloch eigenstates $\ket{u_{n\mathbf{k}}}$ of the unperturbed Hamiltonian,
	\begin{equation}
		H_0(\mathbf{k})\ket{u_{n\mathbf{k}}}=\epsilon_n(\mathbf{k})\ket{u_{n\mathbf{k}}},
		\quad
		\langle u_{m\mathbf{k}}|u_{n\mathbf{k}}\rangle=\delta_{mn}.
		\label{eq:bloch_basis_appendix}
	\end{equation}
	For compactness, we write $\ket{n}\equiv\ket{u_{n\mathbf{k}}}$ below.
	In the Bloch basis, the first-order kinetic equation Eq.~\eqref{eq:rho1_freq_eom} separates naturally into off-diagonal and diagonal sectors. Using $H_{\mathrm E}(\omega_1)=-qE^b(\omega_1)x^b$, one finds
	\begin{equation}
		i(\omega_{mn}-\omega_1)\rho^{(1)}_{mn}(\omega_1)
		=
		\frac{iq}{\hbar}E^b(\omega_1)[x^b,\rho^{(0)}]_{mn},
		\label{eq:rho1_band}
	\end{equation}
	where $\omega_{mn}\equiv(\epsilon_m-\epsilon_n)/\hbar$.
	Using the decomposition $x^b=R^b+r^b$ and $\rho^{(0)}_{mn}=f_n\delta_{mn}$, the off-diagonal and diagonal sectors can be evaluated separately. For $m\neq n$, only the interband position operator contributes:
	\begin{equation}
		[r^b,\rho^{(0)}]_{mn}=
		\sum_{\ell}r^b_{m\ell}\rho^0_{\ell n}-\rho^0_{m\ell}r^b_{\ell n}=
		r^b_{mn}f_{nm},
		\label{eq:xrho_offdiag}
	\end{equation}
	where $f_{nm}\equiv f_n-f_m$, and
	\begin{equation}
		f_n \equiv \frac{1}{e^{(\epsilon_{n\mathbf{k}}-\mu)/(k_B T)}+1}
		\label{eq:fd_dist}
	\end{equation}
	is the Fermi-Dirac distribution function. 
	Substituting Eq.~\eqref{eq:xrho_offdiag} into Eq.~\eqref{eq:rho1_band} gives
	\begin{equation}
		\rho^{(1)}_{mn}(\omega_1)
		=
		-\frac{q}{\hbar}E^b(\omega_1)\frac{r^b_{mn}f_{mn}}{\omega_{mn}-\omega_1},
		\label{eq:rho1_offdiag_appendix}
	\end{equation}
	where we drop the notation $m\neq n$ because $r$ indicates interband only.
	For $m=n$, Eq.~\eqref{eq:rho1_band} becomes
	\begin{equation}
		\rho^{(1)}_{nn}(\omega_1)
		=
		-\frac{iq}{\hbar\omega_1}E^b(\omega_1)\partial_b f_n,
		\label{eq:rho1_diag}
	\end{equation}
	because $[R^b,\rho^{(0)}]_{nn}=i\partial_b\rho^{(0)}_{nn}=i\partial_b f_n$.
	\subsection{second-order density matrix}
	\label{appendix:rho2_length}
	
	We now turn to the second-order density matrix, whose off-diagonal and diagonal sectors must both be retained in the nonlinear current response. Keeping the two ordered input frequencies explicit, and using
	\begin{equation}
		H_{\mathrm E}(\omega_1)=-q\,E^b(\omega_1)x^b,\qquad
		H_{\mathrm E}(\omega_2)=-q\,E^c(\omega_2)x^c,
	\end{equation}
	the frequency-resolved second-order kinetic equation, Eq.~\eqref{eq:rho2_freq_eom}, becomes
	\begin{equation}
		\begin{split}
			i(\omega_{mn}-\Omega)\rho^{(2)}_{mn}(\Omega;\omega_1,\omega_2)
			=&\frac{iq}{\hbar}\Big(
			E^b(\omega_1)[x^b,\rho^{(1)c}(\omega_2)]_{mn}\\
			&+
			E^c(\omega_2)[x^c,\rho^{(1)b}(\omega_1)]_{mn}
			\Big),
		\end{split}
		\label{eq:rho2_band_start}
	\end{equation}
	where the superscript on $\rho^{(1)c}(\omega_2)$ indicates the Cartesian field index carried by the corresponding first-order response. This form makes the interchange symmetry $(b,\omega_1)\leftrightarrow(c,\omega_2)$ manifest. We now follow the tree diagram shown in Fig.~\ref{fig1:rho2} to evaluate each contribution.
	
	\subsubsection{Off-diagonal sector: $m\neq n$}
	
	We first consider the contribution generated by the interband position operator acting on the off-diagonal first-order density matrix [Eq.~\eqref{eq:rho1_offdiag_appendix}]. For a generic input frequency $\omega$, one finds
	\begin{equation}
		\begin{split}
			[r^b,\rho_{\rm off}^{(1)c}(\omega)]_{mn}
			=
			-\frac{q}{\hbar}E^c(\omega)
			&\sum_{\ell\neq m,n}
			\Big(
			r^b_{m\ell}r^c_{\ell n}\frac{f_{\ell n}}{\omega_{\ell n}-\omega}\\
			&-
			r^c_{m\ell}r^b_{\ell n}\frac{f_{m\ell}}{\omega_{m\ell}-\omega}
			\Big).
		\end{split}
		\label{eq:r_rho1off_mn}
	\end{equation}
	The interband position operator acting on the diagonal first-order density matrix, Eq.~\eqref{eq:rho1_diag}, gives
	\begin{equation}
		\begin{split}
			[r^b,\rho^{(1)c}_{\rm diag}(\omega)]_{mn}&=r^b_{mn}\big(\rho^{(1)c}_{nn}-\rho^{(1)c}_{mm}\big)\\
			&=\frac{i q}{\hbar\,\omega}E^c(\omega)\,r^b_{mn}\,\partial_c f_{mn}.
		\end{split}
		\label{eq:r_rho1diag_mn}
	\end{equation}
	
	We next evaluate the contribution generated by the intraband position operator acting on the off-diagonal first-order density matrix:
	\begin{widetext}
		\begin{equation}
			\begin{aligned}
				[R^b,\rho_{\rm off}^{(1)c}(\omega)]_{mn}
				&=
				i\,\rho^{(1)c;b}_{mn}(\omega)
				\\
				&=
				-i\frac{q}{\hbar}E^c(\omega)
				\left[
				\partial_b\!\left(
				r^c_{mn}\frac{f_{mn}}{\omega_{mn}-\omega}
				\right)
				-i\,r^c_{mn}\frac{f_{mn}}{\omega_{mn}-\omega}\,(\mathcal{A}^b_{mm}-\mathcal{A}^b_{nn})
				\right]
				\\
				&=
				-i\frac{q}{\hbar}E^c(\omega)
				\left[
				\frac{f_{mn}}{\omega_{mn}-\omega}\,r^{c;b}_{mn}
				+
				r^c_{mn}\partial_b\!\left(
				\frac{f_{mn}}{\omega_{mn}-\omega}
				\right)
				\right],
			\end{aligned}
			\label{eq:R_rho1off_mn}
		\end{equation}
	\end{widetext}
	where we use Eq.~\eqref{eq:gen_deriv} to simplify the term.
	
	Finally, the intraband position operator acting on the diagonal first-order density matrix does not contribute to the off-diagonal sector:
	\begin{equation}
		[R^b,\rho^{(1)c}_{\rm diag}(\omega)]_{mn}=0,
		\qquad m\neq n.
		\label{eq:R_rho1diag_zero_mn}
	\end{equation}
	Indeed, since $\rho^{(1)}_{\rm diag}$ is band diagonal, its off-diagonal generalized derivative vanishes identically.
	
	Now inserting Eq.~\eqref{eq:r_rho1off_mn}, Eq.~\eqref{eq:r_rho1diag_mn}, and Eq.~\eqref{eq:R_rho1off_mn} into Eq.~\eqref{eq:rho2_band_start}, we obtain the off-diagonal sector:
	\begin{equation}
		\begin{split}
			\rho^{(2)}_{mn}(\Omega)
			&=
			-\frac{q^2E^b(\omega_1)E^c(\omega_2)}{\hbar^2(\omega_{mn}-\Omega)}
			\Big(
			\mathcal{K}^{bc}_{mn}(\omega_2)+\mathcal{K}^{cb}_{mn}(\omega_1)
			\Big),   
		\end{split}
		\label{eq:rho2_off_appendix}
	\end{equation}
	where we have combined the contributions into the kernel
	\begin{equation}
		\begin{split}
			&\mathcal{K}^{bc}_{mn}(\omega)=i\,r^{c;b}_{mn}\frac{f_{mn}}{\omega_{mn}-\omega}
			+i\,r^c_{mn}\partial_b\left(\frac{f_{mn}}{\omega_{mn}-\omega}\right)\\
			&+\sum_{\ell\neq m,n}\Big(
			r^b_{m\ell}r^c_{\ell n}\frac{f_{\ell n}}{\omega_{\ell n}-\omega}
			-r^{c}_{m\ell}r^b_{\ell n}\frac{f_{m\ell}}{\omega_{m\ell}-\omega}
			\Big)\\
			&-\frac{i}{\omega}r^b_{mn}\partial_cf_{mn}.   
		\end{split}    
	\end{equation}

	\subsubsection{Diagonal sector: $m=n$}
	
	We now evaluate the diagonal second-order density matrix. In this sector, the contribution generated by the off-diagonal first-order density matrix is carried by the interband position operator, while the contribution generated by the diagonal first-order density matrix is carried by the intraband position operator, as illustrated in Fig.~\ref{fig1:rho2}.
	Equation~\eqref{eq:rho2_band_start} becomes
	\begin{equation}
		\begin{split}
			-i\Omega\,\rho^{(2)}_{nn}(\Omega;\omega_1,\omega_2)
			=
			\frac{iq}{\hbar}\Big(
			E^b(\omega_1)[x^b,\rho^{(1)c}(\omega_2)]_{nn}\\
			+
			E^c(\omega_2)[x^c,\rho^{(1)b}(\omega_1)]_{nn}
			\Big).
		\end{split}
		\label{eq:rho2_diag_start}
	\end{equation}
	
	We first evaluate the commutator of the interband position operator with the off-diagonal first-order density matrix:
	\begin{equation}
		\begin{split}
			[r^b,\rho_{\rm off}^{(1)c}(\omega)]_{nn}
			&=
			-\frac{q}{\hbar}E^c(\omega)\sum_{\ell\neq n}
			\Bigg[
			r^b_{n\ell}r^c_{\ell n}\frac{f_{\ell n}}{\omega_{\ell n}-\omega}
			\\
			&\hspace{2.6cm}
			-r^c_{n\ell}r^b_{\ell n}\frac{f_{n\ell}}{\omega_{n\ell}-\omega}
			\Bigg]
			\\
			&=
			-\frac{q}{\hbar}E^c(\omega)\sum_{\ell\neq n}f_{\ell n}
			\left(
			\frac{r^b_{n\ell}r^c_{\ell n}}{\omega_{\ell n}-\omega}
			-\frac{r^c_{n\ell}r^b_{\ell n}}{\omega_{\ell n}+\omega}
			\right).
		\end{split}
		\label{eq:r_rho1off_nn}
	\end{equation}
	By contrast, the interband position operator acting on the diagonal first-order density matrix gives no diagonal contribution,
	\begin{equation}
		[r^b,\rho_{\rm diag}^{(1)c}(\omega)]_{nn}=0,
		\label{eq:r_rho1diag_zero_nn}
	\end{equation}
	since $r^b$ is purely off diagonal while $\rho_{\rm diag}^{(1)c}$ is band diagonal.
	
	The intraband position operator acting on the off-diagonal first-order density matrix gives no diagonal contribution,
	\begin{equation}
		[R^b,\rho_{\rm off}^{(1)c}(\omega)]_{nn}=i\left(\partial_b\rho_{\rm off}^{(1)c}(\omega)\right)_{nn}=0,
		\label{eq:R_rho1off_zero_nn}
	\end{equation}
	since $(\rho_{\rm off}^{(1)c})_{nn}=0$.
	The remaining intraband contribution is generated by the diagonal first-order density matrix:
	\begin{equation}
		[R^b,\rho_{\rm diag}^{(1)c}(\omega)]_{nn}
		=
		i\,\partial_b \rho^{(1)c}_{nn}(\omega)
		=
		\frac{q}{\hbar\,\omega}E^c(\omega)\,\partial_b\partial_c f_n.
		\label{eq:R_rho1diag_nn}
	\end{equation}
	Collecting Eqs.~\eqref{eq:r_rho1off_nn} and \eqref{eq:R_rho1diag_nn} and substituting them into Eq.~\eqref{eq:rho2_diag_start}, we obtain
	\begin{equation}
		\begin{split}
			&\rho^{(2)}_{nn}(\Omega;\omega_1,\omega_2)
			=
			\frac{q^2E_b(\omega_1)E_c(\omega_2)}{\hbar^2 \Omega}
			\Bigg[
			-\frac{\partial_b\partial_c f_n}{\omega_2}\\
			&+\sum_{\ell \neq n}f_{\ell n}\left(
			\frac{r^b_{n\ell}r^c_{\ell n}}{\omega_{\ell n}-\omega_2}
			-\frac{r^c_{n\ell}r^b_{\ell n}}{\omega_{\ell n}+\omega_2}
			\right)-\frac{\partial_c\partial_b f_n}{\omega_1}\\
			&+\sum_{\ell \neq n}f_{\ell n}\left(
			\frac{r^c_{n\ell}r^b_{\ell n}}{\omega_{\ell n}-\omega_1}
			-\frac{r^b_{n\ell}r^c_{\ell n}}{\omega_{\ell n}+\omega_1}
			\right)
			\Bigg].
		\end{split}
		\label{eq:rho2_diag_appendix}
	\end{equation}
	
	\section{Position-operator algebra in the length gauge}\label{{appendix: current_in_length}}
	\label{appendix:position_algebra}
	In the length gauge, the position operator is decomposed as
	\begin{equation}
		x^a = R^a + r^a,
	\end{equation}
	where $R^a$ is the intraband part and $r^a$ is the interband part. A central role is played by the covariant derivative of the interband position matrix. In particular, we will use the identities~\cite{Aversa1995}
	\begin{equation}
		\frac{\partial \mathcal{A}^a_{nn}}{\partial k_b}-\frac{\partial \mathcal{A}^b_{nn}}{\partial k_a}
		=
		-i\sum_p\left(r^a_{np}r^b_{pn}-r^b_{np}r^a_{pn}\right)
		=
		-i[r^a,r^b]_{nn},
		\label{eq:AS23}
	\end{equation}
	and
	\begin{equation}
		r^{a;b}_{nm}-r^{b;a}_{nm}
		=
		-i\sum_p\left(r^a_{np}r^b_{pm}-r^b_{np}r^a_{pm}\right)
		=
		-i[r^a,r^b]_{nm}.
		\label{eq:AS24}
	\end{equation}
	The identities in Eqs.~\eqref{eq:AS23} and \eqref{eq:AS24} guarantee that the position operators commute. It is clearest to verify this separately in the off-diagonal and diagonal sectors.
	
	For $m\neq n$, one has
	\begin{equation}
		[x^a,x^b]_{mn}
		=
		[R^a,r^b]_{mn}
		+[r^a,R^b]_{mn}
		+[r^a,r^b]_{mn},
	\end{equation}
	since $[R^a,R^b]_{mn}=0$. Using
	\begin{equation}
		[R^a,r^b]_{mn}=i\,r^{b;a}_{mn},
		\qquad
		[r^a,R^b]_{mn}=-i\,r^{a;b}_{mn},
	\end{equation}
	we obtain
	\begin{equation}
		[x^a,x^b]_{mn}
		=
		-i\big(r^{a;b}_{mn}-r^{b;a}_{mn}\big)+[r^a,r^b]_{mn}.
	\end{equation}
	Equation~\eqref{eq:AS24} then immediately gives $[x^a,x^b]_{mn}=0$.
	
	For $m=n$, the mixed terms vanish and one finds
	\begin{equation}
		[x^a,x^b]_{nn}
		=
		[R^a,R^b]_{nn}
		+[r^a,r^b]_{nn}.
	\end{equation}
	Using
	\begin{equation}
		[R^a,R^b]_{nn}
		=
		i(\partial_a \mathcal{A}^b_{nn}-\partial_b \mathcal{A}^a_{nn}),
	\end{equation}
	together with Eq.~\eqref{eq:AS23}, we obtain
	\begin{equation}
		[R^a,R^b]_{nn}=-[r^a,r^b]_{nn},
	\end{equation}
	and therefore $[x^a,x^b]_{nn}=0$.
	
	\subsection{Proof of Eqs.~\eqref{eq:AS23} and \eqref{eq:AS24}}
	We first prove the diagonal identity, Eq.~\eqref{eq:AS23}. Starting from
	\begin{equation}
		\mathcal{A}^a_{nn}=i\langle n|\partial_a n\rangle,
	\end{equation}
	we compute
	\begin{equation}
		\partial_b \mathcal{A}^a_{nn}-\partial_a \mathcal{A}^b_{nn}
		=
		i\Big(\langle \partial_b n|\partial_a n\rangle-\langle \partial_a n|\partial_b n\rangle\Big),
		\label{eq:diagA_start}
	\end{equation}
	where the mixed second-derivative terms cancel because $\partial_a\partial_b=\partial_b\partial_a$. Inserting completeness, $\sum_p |p\rangle\langle p|=\mathds{1}$, gives
	\begin{equation}
		\begin{split}
			\partial_b \mathcal{A}^a_{nn}-\partial_a \mathcal{A}^b_{nn}
			&=
			i\sum_p\Big(
			\langle \partial_b n|p\rangle\langle p|\partial_a n\rangle \\
			&\qquad\qquad
			-\langle \partial_a n|p\rangle\langle p|\partial_b n\rangle
			\Big).
		\end{split}
		\label{eq:diagA_complete}
	\end{equation}
	Using the orthonormality condition $\langle n|p\rangle=\delta_{np}$, one has
	\begin{equation}
		\partial_a\langle n|p\rangle=0
		\;\Rightarrow\;
		\langle \partial_a n|p\rangle=-\langle n|\partial_a p\rangle.
	\end{equation}
	With the convention $\mathcal{A}^a_{np}=i\langle n|\partial_a p\rangle$, this gives
	\begin{equation}
		\langle n|\partial_a p\rangle=-i\mathcal{A}^a_{np},
		\qquad
		\langle \partial_a n|p\rangle=+i\mathcal{A}^a_{np}.
		\label{eq:diagA_overlap}
	\end{equation}
	Substituting Eq.~\eqref{eq:diagA_overlap} into Eq.~\eqref{eq:diagA_complete}, we obtain
	\begin{equation}
		\begin{split}
			\partial_b \mathcal{A}^a_{nn}-\partial_a \mathcal{A}^b_{nn}
			&=
			-i\sum_p\Big(\mathcal{A}^a_{np}\mathcal{A}^b_{pn}-\mathcal{A}^b_{np}\mathcal{A}^a_{pn}\Big)\\
			&=
			-i\sum_{p\neq n}\Big(r^a_{np}r^b_{pn}-r^b_{np}r^a_{pn}\Big),
		\end{split}
		\label{eq:diagA_final}
	\end{equation}
	where the $p=n$ term vanishes identically, so only interband contributions remain. This proves Eq.~\eqref{eq:AS23}.
	
	We next prove the off-diagonal identity, Eq.~\eqref{eq:AS24}. Starting from the definition of the generalized derivative Eq.~\eqref{eq:r_gen_deriv}, we write
	\begin{equation}
		\begin{split}
			r^{a;b}_{nm}-r^{b;a}_{nm}
			&=
			(\partial_b r^a_{nm}-\partial_a r^b_{nm})
			-i(\mathcal{A}^b_{nn}-\mathcal{A}^b_{mm})r^a_{nm}\\
			&+i(\mathcal{A}^a_{nn}-\mathcal{A}^a_{mm})r^b_{nm}.
		\end{split}
		\label{eq:covdiff_start}
	\end{equation}
	Using $r^a_{nm}=i\langle n|\partial_a m\rangle$, the derivative part becomes
	\begin{equation}
		\partial_b r^a_{nm}-\partial_a r^b_{nm}
		=
		i\Big(\langle \partial_b n|\partial_a m\rangle-\langle \partial_a n|\partial_b m\rangle\Big),
		\label{eq:rgendiff}
	\end{equation}
	where the mixed second-derivative terms cancel because $\partial_a\partial_b=\partial_b\partial_a$. Inserting completeness, $\sum_p|p\rangle\langle p|=\mathds{1}$, Eq.~\eqref{eq:rgendiff} can be simplified to
	\begin{equation}
		\begin{split}
			\partial_b r^a_{nm}-\partial_a r^b_{nm}=&
			i\sum_p\Big(
			\langle \partial_b n|p\rangle\langle p|\partial_a m\rangle\\
			&-\langle \partial_a n|p\rangle\langle p|\partial_b m\rangle
			\Big).
		\end{split}
		\label{eq:rgen_overlap}
	\end{equation}
	Substituting the overlap relations Eq.~\eqref{eq:diagA_overlap} into Eq.~\eqref{eq:rgen_overlap}, we get
	\begin{align}
		\partial_b r^a_{nm}-\partial_a r^b_{nm}
		&=
		-i\sum_p\Big(\mathcal{A}^a_{np}\mathcal{A}^b_{pm}-\mathcal{A}^b_{np}\mathcal{A}^a_{pm}\Big)\notag\\
		&=
		-i\sum_{p\neq m,n}\Big(r^a_{np}r^b_{pm}-r^b_{np}r^a_{pm}\Big)\notag\\
		&\quad
		-i(\mathcal{A}^a_{nn}-\mathcal{A}^a_{mm})r^b_{nm}\notag\\
		&\quad
		+i(\mathcal{A}^b_{nn}-\mathcal{A}^b_{mm})r^a_{nm},
		\label{eq:rawdiff_final}
	\end{align}
	where we write down $p=m$ and $p=n$ terms explicitly in the second line.
	Substituting Eq.~\eqref{eq:rawdiff_final} into Eq.~\eqref{eq:covdiff_start}, the Berry-connection terms cancel, leaving
	\begin{equation}
		r^{a;b}_{nm}-r^{b;a}_{nm}
		=
		-i\sum_p\Big(r^a_{np}r^b_{pm}-r^b_{np}r^a_{pm}\Big),
		\label{eq:AS24_proof}
	\end{equation}
	which proves Eq.~\eqref{eq:AS24}.
	
	\subsection{Relation to intraband and interband current decompositions} \label{append:intra}
	We have established that the total current operator in the length gauge does not acquire an explicit field-dependent correction, because
	\begin{equation}
		v^a=\frac{i}{\hbar}[H,x^a]=\frac{i}{\hbar}[H_0,x^a],
	\end{equation}
	with $H=H_0+H_{\mathrm E}$ and $[H_{\mathrm E},x^a]=0$. Thus the electric field enters the total nonlinear current only through the perturbative density matrix.
	
	This should be distinguished from formulations that decompose the response into intraband and interband polarization channels~\cite{Aversa1995,Sipe2000}. Writing
	\begin{equation}
		\begin{aligned}
			P^a(t)&=P^a_{\mathrm{intra}}(t)+P^a_{\mathrm{inter}}(t),\\
			P^a_{\mathrm{intra}}&=q\,\mathrm{Tr}[R^a\rho],\\
			P^a_{\mathrm{inter}}&=q\,\mathrm{Tr}[r^a\rho].
		\end{aligned}
		\label{eq:P_split}
	\end{equation}
	the corresponding channel-resolved currents are defined by
	\begin{equation}
		\begin{aligned}
			J^a_{\mathrm{intra}}(t)
			&=\dv{P^a_{\mathrm{intra}}}{t}
			=\frac{iq}{\hbar}\,\mathrm{Tr}\!\left(\rho[H,R^a]\right),\\
			J^a_{\mathrm{inter}}(t)
			&=\dv{P^a_{\mathrm{inter}}}{t}
			=\frac{iq}{\hbar}\,\mathrm{Tr}\!\left(\rho[H,r^a]\right).
		\end{aligned}
		\label{eq:J_split_def}
	\end{equation}
	Because $R^a$ and $r^a$ do not separately commute with the perturbation, one finds
	\begin{equation}
		\begin{aligned}
			&[H,R^a]=[H_0,R^a]+[H_{\mathrm E},R^a]=[H_0,R^a]-qE^b[x^b,R^a], \\ 
			&[H,r^a]=[H_0,r^a]+[H_{\mathrm E},r^a]=[H_0,r^a]-qE^b[x^b,r^a].
		\end{aligned}
		\label{eq:H_split_comm}
	\end{equation}
	These additional commutator terms are absent in the total current because they cancel when $R^a$ and $r^a$ are recombined into $x^a=R^a+r^a$, but they survive in the channel-resolved currents and appear as the effective velocity corrections discussed in the intraband/interband decomposition of Aversa and Sipe~\cite{Aversa1995,Sipe2000}.

	\section{Conductivity tensor in length gauge}\label{appendix:conductivity_LG}
	\subsection{Linear order response}\label{appendix:linear_response}
	At linear order, the frequency-matching condition is implicit, and the current takes the form
	\begin{equation}
		\begin{aligned}
			&J^{a,(1)}(t)=\int\frac{d\omega}{2\pi}\,J^{a,(1)}(\omega)e^{-i\omega t},\\
			&J^{a,(1)}(\omega)=\sigma^{ab}(\omega)E^b(\omega).    
		\end{aligned}
		\label{eq:J1_sigma}
	\end{equation}
	Substituting Eq.~\eqref{eq:rho1_main} into Eq.~\eqref{eq:J_n_def}, one obtains
	\begin{equation}
		\begin{split}
			\sigma^{ab}(\omega)
			&=
			-\frac{q^2}{\hbar}\int[dk]\Bigg(
			\sum_n v^a_{nn}\frac{i\,\partial_b f_n}{\omega}\\
			&+
			\sum_{n,m\neq n}v^a_{nm}r^b_{mn}\frac{f_{mn}}{\omega_{mn}-\omega}
			\Bigg),
		\end{split}
		\label{eq:sigma_ab}
	\end{equation}
	where the first term is the intraband contribution from the diagonal part of $\rho^{(1)}$, while the second term arises from the interband coherence. For notational simplicity, we suppress the Brillouin-zone integral $\int[dk]$ in the following expressions. 
	
	The conductivity in Eq.~\eqref{eq:sigma_ab} is obtained first at finite frequency. The strict dc limit is then defined by the adiabatic prescription $\omega\to i\eta$ with $\eta\to0^+$, consistent with our Fourier convention $e^{-i\omega t}$~\cite{Das2023}. Applying this to Eq.~\eqref{eq:sigma_ab} gives
	\begin{equation}
		\sigma^{ab}_{\rm dc}
		=
		-\frac{q^2}{\hbar}\Bigg(
		\sum_n v^a_{nn}\frac{\partial_b f_n}{\eta}
		+
		\sum_{m\neq n}v^a_{nm}r^b_{mn}\frac{f_{mn}}{\omega_{mn}-i\eta}
		\Bigg).
		\label{eq:sigma_dc_split}
	\end{equation}
	Using $v^a_{nn}=\hbar^{-1}\partial_a\epsilon_n$, the first term becomes
	\begin{equation}
		\sigma^{ab}_{\rm Drude}
		=
		-\frac{q^2}{\hbar^2\eta}\sum_n (\partial_a\epsilon_n)(\partial_b f_n),
		\label{eq:sigma_drude}
	\end{equation}
	which is the usual clean-limit Drude contribution. By contrast, the second term remains finite as $\eta\to0^+$. Using $v^a_{nm}=i\omega_{nm}r^a_{nm}$ for $m\neq n$, one finds the anomalous Hall conductivity~\cite{Nagaosa2010}
	\begin{equation}
		\begin{split}
			\sigma^{ab}_{\rm inter}
			&=-\frac{i q^2}{\hbar}\sum_{n,m\neq n} r^a_{nm}r^b_{mn}f_{nm},\\
			&=-\frac{i q^2}{\hbar}\sum_{n}f_n\sum_{m\neq n}(r^a_{nm}r^b_{mn}-r^b_{nm}r^a_{mn})\\
			&=-\frac{q^2}{\hbar}\sum_nf_n\Omega^{ab}_n,
		\end{split}
		\label{eq:sigma_inter_dc}
	\end{equation}
	where $\Omega^{ab}_n\equiv i\sum_{m\neq n}(r^a_{nm}r^b_{mn}-r^b_{nm}r^a_{mn})$ is the Berry curvature~\cite{Niu2010}.
	
	\subsection{second-order response}\label{appendix:2nd_response}
	We now analyze the $\eta$-independent contributions from the diagonal [Eq.~\eqref{eq:chi_diag}] and off-diagonal [Eq.~\eqref{eq:chi_od}] sectors of the nonlinear conductivity tensor in the adiabatic dc limit. In the following few equations, we suppress the common overall prefactors \(q^3/\hbar^2\) for brevity.

	\subsubsection{Diagonal sector}
	
	In Eq.~\eqref{eq:chi_diag}, the $\partial_b\partial_cf_n/(\omega_2\Omega)$ term diverges as $\eta^{-2}$ and is a nonlinear Drude-like term~\cite{Das2023,Su2024,Jia2024}. The remaining diagonal contribution, however, survives in the clean limit.
	\begin{widetext}
		\begin{equation}
			\begin{split}
				\chi^{abc}_{\rm diag}(2i\eta;i\eta,i\eta)&=\frac{1}{ 2i\eta}\sum_{n,m \neq n}v^a_{nn}f_{mn}\left(
				\frac{r^b_{nm}r^c_{mn}}{\omega_{m n}-i\eta}
				-\frac{r^c_{nm}r^b_{mn}}{\omega_{mn}+i\eta}
				+
				\frac{r^c_{nm}r^b_{mn}}{\omega_{mn}-i\eta}
				-\frac{r^b_{nm}r^c_{mn}}{\omega_{mn}+i\eta}
				\right)\\
				&=\frac{1}{2i\eta}\sum_{n,m \neq n}v^a_{nn}f_{mn}\left(
				\frac{r^b_{nm}r^c_{mn}2i\eta}{\omega^2_{m n}+\eta^2}
				+\frac{r^c_{nm}r^b_{mn}2i\eta}{\omega^2_{mn}+\eta^2}
				\right)\\
				&=\sum_{n,m \neq n}v^a_{nn}f_{mn}
				\frac{r^b_{nm}r^c_{mn}+r^c_{nm}r^b_{mn}}{\omega^2_{m n}+\eta^2}.
			\end{split}
		\end{equation}
	\end{widetext}
	In the limit of $\eta\to0^+$, the surviving term is denoted as 
	\begin{equation}
		\begin{split}
			\chi^{abc}_{\rm dc, diag}&\equiv\sum_{n,m \neq n}2f_{mn}g^{bc}_{nm}\frac{v^a_{nn}}{\omega^2_{m n}}\\
			&=\sum_{n,m \neq n}f_{mn}g^{bc}_{nm}\frac{v^a_{nn}-v^a_{mm}}{\omega^2_{m n}}\\
			&=\sum_{n,m \neq n}f_{mn}g^{bc}_{nm}\frac{\Delta^a_{nm}}{\omega^2_{m n}},
		\end{split}
		\label{eq:dc_diag_appendix}
	\end{equation}
	where in the second line we symmetrize the double sum under $m\leftrightarrow n$, $g^{bc}_{nm}\equiv \frac{1}{2}( r^b_{nm}r^c_{mn}+r^c_{nm}r^b_{mn})$ is the quantum metric tensor, and $\Delta^a_{nm}\equiv v^a_{nn}-v^a_{mm}$.
	Although Eq.~\eqref{eq:dc_diag_appendix} appears at first sight to be a pure Fermi sea term, it actually contains both Fermi sea and Fermi surface pieces. Using the identity
	\begin{equation}
		\frac{\Delta^a_{nm}}{\omega^2_{m n}}=-\partial_a(\frac{1}{\omega_{nm}}),
		\label{eq:Delta}
	\end{equation}
	we rewrite Eq.~\eqref{eq:dc_diag_appendix} as
	\begin{equation}
		\begin{split}
			\chi^{abc}_{\rm diag}
			&=
			-\sum_{n,m \neq n}f_{mn}g^{bc}_{nm}\partial_a(\frac{1}{\omega_{nm}})
			\\
			&=-\sum_{n,m \neq n}\Bigg(\partial_a\Big(\frac{f_{mn}g^{bc}_{nm}}{\omega_{nm}}\Big)-\frac{\partial_a\Big(f_{mn}g^{bc}_{nm}\Big)}{\omega_{nm}}\Bigg)\\
			&=
			\sum_{n,m\neq n}\frac{g^{bc}_{nm}\partial_af_{mn}}{\omega_{nm}}+\frac{f_{mn}\partial_ag^{bc}_{nm}}{\omega_{nm}}
			\\
			&=
			\sum_{n,m\neq n}\frac{g^{bc}_{nm}\partial_af_{mn}}{\omega_{nm}}-F_{mn}
			\big(r^{b;a}_{nm}r^c_{mn}+r^b_{nm}r^{c;a}_{mn}).
			\\
		\end{split}
		\label{eq:chi_diag_static_ibp}
	\end{equation}
	where we define $F_{mn}=\frac{f_{mn}}{\omega_{mn}}=F_{nm}$ for brevity. In going from the first to the second line, we integrate by parts over the Brillouin zone and drop the boundary term.
	In going from the third to the fourth line, we use the product rule together with the definition of the generalized derivative:
	\begin{equation}
		\partial_a(r^b_{nm}r^c_{mn})=r^{b;a}_{nm}r^c_{mn}+r^b_{nm}r^{c;a}_{mn}.
		\label{eq:genr_prod}
	\end{equation}
	Here the Berry-connection terms proportional to $(\mathcal{A}^a_{nn}-\mathcal{A}^a_{mm})$ cancel between the two factors. 
	The first term is explicitly of Fermi surface type, since it is proportional to $\partial_a f_{mn}$ and therefore vanishes in an insulator. By contrast, the second term appears at this stage to define a finite and gauge-invariant Fermi sea contribution that can survive in an insulator. However, this interpretation would be premature: additional Fermi sea–like terms arise from the off-diagonal sector, and only the sum of all such terms has physical meaning. As we show below, these apparent Fermi sea contributions cancel exactly in the strict adiabatic dc limit.
	
	\subsubsection{Off-diagonal sector and cancellation of apparent Fermi sea terms}
	We insert the kernel in Eq.~\eqref{eq:Kappa} into Eq.~\eqref{eq:chi_od} and analyze the adiabatic static limit by replacing $\omega\to i\eta$ and $\Omega\to 2i\eta$:
	\begin{equation}
		\begin{split}
			&\chi^{abc}_{\rm od}(2i\eta;i\eta,i\eta)=
			\sum_{n,m\neq n}
			\frac{-i\omega_{nm}r^a_{nm}}{\omega_{mn}-2i\eta}
			\Bigg[-\frac{1}{\eta}r^b_{mn}\partial_cf_{mn}\\
			&+i\,r^{c;b}_{mn}\frac{f_{mn}}{\omega_{mn}-i\eta}+i\,r^c_{mn}\partial_b\left(\frac{f_{mn}}{\omega_{mn}-i\eta}\right)\\
			&+\sum_{\ell\neq m,n}\Big(
			r^b_{m\ell}r^c_{\ell n}\frac{f_{\ell n}}{\omega_{\ell n}-i\eta}
			-r^{c}_{m\ell}r^b_{\ell n}\frac{f_{m\ell}}{\omega_{m\ell}-i\eta}
			\Big)+b \leftrightarrow c  
			\Bigg].
		\end{split}
		\label{eq:chiod_static}
	\end{equation}
	Note that we use $v^a_{nm}=i\omega_{nm}r^a_{nm}$ to express the off-diagonal contribution entirely in terms of interband position matrix elements.
	
	To extract the intrinsic dc contribution, we keep only the $\eta^0$ terms as $\eta\to0^+$. The explicit Fermi surface term proportional to $\partial_c f_{mn}$ requires a separate expansion of the prefactor \(\frac{-i\omega_{nm}}{\omega_{mn}-2i\eta}\), while in the remaining terms one may set $\eta\to0$ directly.
	Using
	\begin{equation}
		\frac{-i\omega_{nm}}{\omega_{mn}-2i\eta}
		=
		i-\frac{2\eta}{\omega_{mn}-2i\eta},
		\label{eq:prefactor_expand}
	\end{equation}
	the term proportional to \(\partial_c f_{mn}\) gives
	\begin{equation}
		\left(\frac{-i\omega_{nm}}{\omega_{mn}-2i\eta}\right)\left(-\frac{1}{\eta}\right)
		=
		-\frac{i}{\eta}
		+\frac{2}{\omega_{mn}-2i\eta},
		\label{eq:prefactor_expand_dc}
	\end{equation}
	so that its $\eta$-independent part, which is necessarily of Fermi surface type, is
	\begin{equation}
		\chi^{abc}_{\rm od,sur}
		\equiv
		\sum_{n,m\neq n}\frac{2}{\omega_{mn}}
		\Big(r^a_{nm}r^b_{mn}\partial_c f_{mn}+r^a_{nm}r^c_{mn}\partial_b f_{mn}\Big).
		\label{eq:od_sur_raw}
	\end{equation}
	Using the symmetry of the double sum under $m\leftrightarrow n$, this may be rewritten as
	\begin{equation}
		\chi^{abc}_{\rm od,sur}
		=
		\sum_{n,m\neq n}\frac{2}{\omega_{mn}}
		\Big(g^{ab}_{nm}\partial_c f_{mn}+g^{ca}_{nm}\partial_b f_{mn}\Big),
		\label{eq:od_sur}
	\end{equation}
	where
	\begin{equation}
		g^{ab}_{nm}\equiv \frac12\left(r^a_{nm}r^b_{mn}+r^b_{nm}r^a_{mn}\right).
	\end{equation}
	The remaining $\eta$-independent Fermi sea part becomes
	\begin{equation}
		\begin{split}
			&\chi^{abc}_{\rm od,sea}=\sum_{n,m\neq n}\Bigg[-\,r^a_{nm}r^{c;b}_{mn}F_{mn}-r^a_{nm}r^c_{mn}\partial_bF_{mn}\\
			&+\sum_{\ell\neq m,n}ir^a_{nm}\Big(
			r^b_{m\ell}r^c_{\ell n}F_{\ell n}
			-r^{c}_{m\ell}r^b_{\ell n}F_{m\ell}
			\Big)+b \leftrightarrow c  
			\Bigg].
		\end{split}
		\label{eq:chiod_sea}
	\end{equation}
	
	The second term in Eq.~\eqref{eq:chiod_sea} may be integrated by parts and rewritten in terms of generalized derivatives. 
	\begin{equation}
		\begin{split}
			&\sum_{n,m\neq n}-\,r^a_{nm}r^{c;b}_{mn}F_{mn}-r^a_{nm}r^c_{mn}\partial_bF_{mn}+b\leftrightarrow c\\
			&=\sum_{n,m\neq n}-\,r^a_{nm}r^{c;b}_{mn}F_{mn}+F_{mn}\partial_b\big(r^a_{nm}r^c_{mn}\big)+b\leftrightarrow c\\
			&=\sum_{n,m\neq n}F_{mn}(-\,r^a_{nm}r^{c;b}_{mn}+r^{a;b}_{nm}r^c_{mn}+r^a_{nm}r^{c;b}_{mn})+b\leftrightarrow c\\
			&=\sum_{n,m\neq n}F_{mn}r^{a;b}_{nm}r^c_{mn}+F_{mn}r^{a;c}_{nm}r^b_{mn}
		\end{split}
		\label{eq:od_sea_ibp}
	\end{equation}
	
	Combining Eq.~\eqref{eq:od_sea_ibp} with the apparent Fermi sea term from the diagonal sector, we obtain
	\begin{equation}
		\sum_{n,m\neq n}F_{mn}\Big[
		r^c_{mn}\big(r^{a;b}_{nm}-r^{b;a}_{nm}\big)
		+r^b_{mn}\big(r^{a;c}_{nm}-r^{c;a}_{nm}\big)
		\Big].
		\label{eq:sea_combined_commutator_form}
	\end{equation}
	At this stage they appear to define a finite Fermi sea contribution. However, this interpretation is still premature, because the remaining triple-$r$ term in Eq.~\eqref{eq:chiod_sea} can be rearranged and shown to cancel the full apparent Fermi sea part.

	The triple-$r$ contribution may be rewritten as
	\begin{align}
		\chi^{abc}_{\mathrm{od},r}
		&=
		\sum_{\ell,m,n} i\Big(
		r^a_{nm}r^b_{m\ell}r^c_{\ell n}F_{\ell n}
		-r^a_{nm}r^{c}_{m\ell}r^b_{\ell n}F_{m\ell}
		+b\leftrightarrow c
		\Big)
		\notag\\
		&=
		\sum_{\ell,m,n} i\Big(
		r^c_{mn}F_{mn}r^a_{n\ell}r^b_{\ell m}
		-r^{c}_{mn}F_{mn}r^a_{\ell m}r^b_{n\ell}
		+b\leftrightarrow c
		\Big)
		\notag\\
		&=
		\sum_{m,n} F_{mn}\Big(
		r^c_{mn}\,i[r^a,r^b]_{nm}
		+r^b_{mn}\,i[r^a,r^c]_{nm}
		\Big)
		\notag\\
		&=
		\sum_{m,n} -F_{mn}\Big[
		r^c_{mn}\big(r^{a;b}_{nm}-r^{b;a}_{nm}\big)
		+r^b_{mn}\big(r^{a;c}_{nm}-r^{c;a}_{nm}\big)
		\Big].
		\label{eq:triple_r_rewrite}
	\end{align}
	Here the summation over the intermediate band index $\ell$ is restricted to $\ell\neq m,n$, since the interband matrix elements $r^a_{mn}$ vanish when the two band indices are equal.
	In going from the first to the second line, we relabel dummy indices as $(n,m,\ell)\rightarrow(n,\ell,m)$ in the first term and $(n,m,\ell)\rightarrow(\ell,m,n)$ in the second term. The last line then follows from the sum rule in Eq.~\eqref{eq:AS24}.
	
	Equation~\eqref{eq:triple_r_rewrite} exactly cancels the apparent Fermi sea contribution from the generalized-derivative terms in Eq.~\eqref{eq:sea_combined_commutator_form}, so that no Fermi sea term survives in the strict adiabatic dc limit.
	
	\subsubsection{Static conductivity in terms of the BCP tensor}
	Now putting the nonzero Fermi surface terms from diagonal Eq.~\eqref{eq:chi_diag_static_ibp} and off-diagonal Eq.~\eqref{eq:od_sur} sectors together, we achieve our final expression
	\begin{equation}
		\begin{split}
			\chi^{abc}&=\sum_{n,m\neq n}\frac{g^{bc}_{nm}\partial_af_{mn}}{\omega_{nm}}+\frac{2g^{ab}_{nm}\partial_c f_{mn}+2g^{ac}_{nm}\partial_b f_{mn}}{\omega_{mn}}\\
			&=2\sum_{n,m\neq n}-\frac{g^{bc}_{nm}\partial_af_{n}}{\omega_{nm}}+\frac{2g^{ab}_{nm}\partial_c f_{n}+2g^{ac}_{nm}\partial_b f_{n}}{\omega_{nm}}\\
			&=\hbar\sum_{n}-G^{bc}_n\partial_af_{n}+2G^{ab}_n\partial_cf_{n}+2G^{ca}_n\partial_bf_{n},\\    
		\end{split}    
		\label{eq:fermi surface length}
	\end{equation}
	where
	\begin{equation}
		G^{ab}_n\equiv\sum_{m\neq n}\frac{2g^{ab}_{nm}}{\epsilon_{nm}}
		=
		2\,\mathrm{Re}\sum_{m\neq n}\frac{r^a_{nm}r^b_{mn}}{\epsilon_{nm}}
		\label{eq:BCP_def}
	\end{equation}
	is the Berry-connection polarizability tensor~\cite{Gao2014,Xiao2021,Yang2021}.
	
	To compare with the static nonlinear conductivity written directly in terms of the real electric field~\cite{Das2023,Jia2024}, one must convert from Fourier amplitudes to the real-field amplitude.

	For a real field,
	\begin{equation}
		E^b(-\omega)=E^b(\omega)^*,
	\end{equation}
	so that in the static limit
	\begin{equation}
		E^b_{\rm dc}=E^b(\omega)+E^b(-\omega)=2\,\mathrm{Re}\,E^b(\omega),
	\end{equation}
	and hence
	\begin{equation}
		E^b(\omega)\xrightarrow[\omega\to0]{}\frac{E^b_{\rm dc}}{2}.
		\label{eq:E_dc_half}
	\end{equation}
	Similarly, the physical real current is
	\begin{equation}
		J^a_{\rm phys}=J^a(\Omega)+J^a(\Omega)^*=2\,\mathrm{Re}\,J^a(\Omega).
	\end{equation}
	If the static coefficient is real, then
	\begin{equation}
		J^a_{\rm phys}
		=
		2\,\chi^{abc}
		\left(\frac{E^b_{\rm dc}}{2}\right)
		\left(\frac{E^c_{\rm dc}}{2}\right)
		=
		\frac{1}{2}\chi^{abc}E^b_{\rm dc}E^c_{\rm dc}.
	\end{equation}
	We therefore define the static nonlinear conductivity in the real-field convention as
	\begin{equation}
		\chi^{abc}_{\rm dc}\equiv \frac{1}{2}\chi^{abc}=\hbar\sum_{n}G^{ab}_n\partial_cf_{n}+G^{ca}_n\partial_bf_{n}-\frac{1}{2}G^{bc}_n\partial_af_{n},
		\label{eq:chi_dc_real}
	\end{equation}
	and use $\chi^{abc}_{\rm dc}$ in the main text and in comparisons with the literature.
	
	\subsubsection{Dissipative and reactive parts of the static response}
	
	We now decompose the static nonlinear conductivity into dissipative and reactive parts~\cite{Souza2022}. The dc quadratic current is
	\begin{equation}
		J^a_{\rm dc}=\chi^{abc}_{\rm dc}E^bE^c,
	\end{equation}
	and the corresponding power density is
	\begin{equation}
		\dot Q=J^a_{\rm dc}E^a=\chi^{abc}_{\rm dc}E^aE^bE^c,
	\end{equation}
	where repeated Cartesian indices are summed over.
	Since the product $E^aE^bE^c$ is fully symmetric in the Cartesian indices, only the fully symmetrized part of $\chi^{abc}_{\rm dc}$ contributes to dissipation. We therefore define
	\begin{equation}
		\chi^{abc}_{\rm diss}
		\equiv
		\frac{1}{6}\Big(
		\chi^{abc}_{\rm dc}
		+\chi^{acb}_{\rm dc}
		+\chi^{bac}_{\rm dc}
		+\chi^{bca}_{\rm dc}
		+\chi^{cab}_{\rm dc}
		+\chi^{cba}_{\rm dc}
		\Big),
		\label{eq:chi_diss}
	\end{equation}
	so that
	\begin{equation}
		\dot Q=\chi^{abc}_{\rm diss}E^aE^bE^c.
	\end{equation}
	The reactive part is the remainder,
	\begin{equation}
		\chi^{abc}_{\rm rea}
		\equiv
		\chi^{abc}_{\rm dc}-\chi^{abc}_{\rm diss},
		\label{eq:chi_rea}
	\end{equation}
	which satisfies
	\begin{equation}
		\chi^{abc}_{\rm rea}E^aE^bE^c=0.
	\end{equation}
	
	Using the fully symmetrized definition of the dissipative part and the symmetry \(G_n^{ab}=G_n^{ba}\), the physical dc conductivity decomposes as
	\begin{equation}
		\begin{aligned}
			&\chi^{abc}_{\rm dc}=\chi^{abc}_{\rm diss}+  \chi^{abc}_{\rm rea}\\
			&\chi^{abc}_{\rm diss}=\frac{\hbar}{2}\sum_{n}G^{ab}_n\partial_cf_{n}+G^{ca}_n\partial_b f_{n}+G^{bc}_n\partial_a f_{n},\\
			&\chi^{abc}_{\rm rea}=\frac{\hbar}{2}\sum_{n}G^{ab}_n\partial_c f_{n}+G^{ca}_n\partial_b f_{n}-2G^{bc}_n\partial_a f_{n}.\
		\end{aligned}
	\end{equation}
	
	\section{Density matrix in the velocity gauge}\label{appendix:rho_vg}
	
	In this Appendix we provide the details of the velocity gauge derivation of
	the nonlinear conductivity. The purpose is to show explicitly how the
	field-dependent current operators in the velocity gauge combine with the density-matrix response to reproduce the length gauge result. We keep the carrier charge \(q\) arbitrary and use the same Fourier convention
	\[
	X(t)=\int\frac{d\omega}{2\pi}X(\omega)e^{-i\omega t},
	\qquad
	\mathbf{E}(\omega)=i\omega\mathbf{A}(\omega).
	\]
	Throughout this Appendix the current direction is denoted by \(a\), while
	the two perturbing fields carry indices \(b\) and \(c\).
	
	The density matrix obeys the same Liouville equation used in the
	length gauge derivation,
	\begin{equation}
		\dv{\rho}{t}
		=
		-\frac{i}{\hbar}[H_0+H_A(t),\rho],
		\label{eq:Liouville_full_vg}
	\end{equation}
	where in the velocity gauge
	\begin{equation}
		H_A(t)
		=
		-qA^b(t)v^b
		+\frac{q^2}{2}A^b(t)A^c(t)w^{bc}
		+\cdots .
	\end{equation}
	Thus the difference between the two gauges does not come from the
	Liouville equation itself, but from how the external field enters the
	Hamiltonian and current operator.
	
	\subsection{First-order density matrix}
	To first order in the vector potential,
	\begin{equation}
		\dv{\rho^{(1)}}{t}
		=
		-\frac{i}{\hbar}[H_0,\rho^{(1)}]
		-\frac{i}{\hbar}[H_A^{(1)}(t),\rho^{(0)}],
	\end{equation}
	with
	\begin{equation}
		H_A^{(1)}(t)=-qA^b(t)v^b .
	\end{equation}
	In the band basis this gives
	\begin{equation}
		i(\omega_{mn}-\omega_1)\rho^{(1)}_{mn}(\omega_1)
		=
		\frac{iq}{\hbar}A^b(\omega_1)
		[v^b,\rho^{(0)}]_{mn}.
		\label{eq:rho1_vg_band}
	\end{equation}
	Using
	\begin{equation}
		[v^b,\rho^{(0)}]_{mn}
		=
		-f_{mn}v^b_{mn},
	\end{equation}
	we obtain
	\begin{equation}
		\rho^{(1)}_{mn}(\omega_1)
		=
		-\frac{q}{\hbar}A^b(\omega_1)
		\frac{f_{mn}v^b_{mn}}{\omega_{mn}-\omega_1},
		\qquad m\neq n .
		\label{eq:rho1_vg}
	\end{equation}
	Equation~\eqref{eq:rho1_vg} shows an important distinction between the two
	gauges. In the velocity gauge, the first-order density matrix has no
	diagonal component,
	\begin{equation}
		\rho^{(1)}_{nn}(\omega_1)=0,
	\end{equation}
	because \(f_{nn}=0\). Thus the Fermi surface part of the linear response is
	not carried by \(\rho^{(1)}_{nn}\), as in the length gauge, but instead by
	the field-dependent current vertex \(j^{(1)}\rho^{(0)}\).
	
	For the off-diagonal component, using
	\begin{equation}
		v^b_{mn}=i\omega_{mn}r^b_{mn},\qquad m\neq n,
	\end{equation}
	and \(E^b(\omega_1)=i\omega_1 A^b(\omega_1)\), Eq.~\eqref{eq:rho1_vg}
	can be written as
	\begin{equation}
		\rho^{(1)}_{mn}(\omega_1)
		=
		-\frac{qE^b(\omega_1)}{\hbar}
		\frac{\omega_{mn}}{\omega_1}
		\frac{f_{mn}r^b_{mn}}{\omega_{mn}-\omega_1},
		\quad m\neq n .
		\label{eq:rho1_vg_r}
	\end{equation}
	This expression has the same interband matrix element and resonance
	denominator as the length gauge off-diagonal density matrix in
	Eq.~\eqref{eq:rho1_offdiag_appendix}, but it is not identical at finite
	frequency because of the additional factor \(\omega_{mn}/\omega_1\).
	Equation~\eqref{eq:rho1_vg_r} should therefore be compared with
	Eq.~\eqref{eq:rho1_offdiag_appendix} only after the static or
	frequency-independent limit is taken consistently. In that limit, the
	off-diagonal interband part reduces to the corresponding length gauge
	expression, whereas the diagonal Fermi-surface contribution, absent from
	\(\rho^{(1)}\) in the velocity gauge, reappears through the
	\(j^{(1)}\rho^{(0)}\) contribution to the current. Therefore, the equivalence
	with the length gauge response is recovered only after the full velocity gauge
	current, \(j^{(0)}\rho^{(1)}+j^{(1)}\rho^{(0)}\), is included.

	\subsection{second-order density matrix}
	
	To second-order in the vector potential,
	\begin{equation}
		\dv{\rho^{(2)}}{t}
		=
		-\frac{i}{\hbar}[H_0,\rho^{(2)}]
		-\frac{i}{\hbar}[H_A^{(1)}(t),\rho^{(1)}]
		-\frac{i}{\hbar}[H_A^{(2)}(t),\rho^{(0)}],
	\end{equation}
	with
	\begin{equation}
		H_A^{(1)}(t)=-qA^b(t)v^b,
		\qquad
		H_A^{(2)}(t)=\frac{q^2}{2}A^b(t)A^c(t)w^{bc}.
	\end{equation}
	We first keep the ordered contribution proportional to
	\(A^b(\omega_1)A^c(\omega_2)\). The contribution with
	\((b,\omega_1)\leftrightarrow(c,\omega_2)\) is included when constructing
	the symmetrized response.
	
	In the band basis, the frequency-resolved equation becomes
	\begin{equation}
		\begin{split}
			&i(\omega_{mn}-\Omega)\rho^{(2)}_{mn}(\Omega;\omega_1,\omega_2)
			=
			\frac{iq}{\hbar}A^b(\omega_1)
			\comm{v^b}{\rho^{(1)c}(\omega_2)}_{mn}  \\
			&-\frac{iq^2}{2\hbar}
			A^b(\omega_1)A^c(\omega_2)
			\comm{w^{bc}}{\rho^{(0)}}_{mn} +\,(b,\omega_1)\leftrightarrow(c,\omega_2),
		\end{split}
		\label{eq:rho2_vg_band}
	\end{equation}
	where \(\Omega=\omega_1+\omega_2\), and
	\(\rho^{(1)c}(\omega_2)\) denotes the first-order density matrix driven by
	the perturbation \(A^c(\omega_2)\). The final term means that the entire
	ordered contribution on the right-hand side is repeated with
	\((b,\omega_1)\) and \((c,\omega_2)\) interchanged.
	
	For compactness, in the following algebra we focus on the ordered
	\((b,\omega_1;c,\omega_2)\) contribution and write
	\begin{equation}
		A^b\equiv A^b(\omega_1),\quad
		A^c\equiv A^c(\omega_2),\quad
		\rho^{(1)c}\equiv \rho^{(1)c}(\omega_2).
	\end{equation}
	The interchanged contribution is obtained from the final ordered expression
	by applying \((b,\omega_1)\leftrightarrow(c,\omega_2)\).
	We first evaluate the two commutators appearing in
	Eq.~\eqref{eq:rho2_vg_band}. For the equilibrium density matrix,
	\begin{equation}
		\comm{w^{bc}}{\rho^{(0)}}_{mn}
		=
		w^{bc}_{mn}f_n-f_m w^{bc}_{mn}
		=f_{nm}w^{bc}_{mn}.
		\label{eq:w_rho0_comm_vg}
	\end{equation}
	For the first term, we use
	\begin{equation}
		\comm{v^b}{\rho^{(1)c}}_{mn}
		=
		\sum_{\ell}
		\left[
		v^b_{m\ell}\rho^{(1)c}_{\ell n}
		-
		\rho^{(1)c}_{m\ell}v^b_{\ell n}
		\right].
		\label{eq:v_rho1_comm_vg}
	\end{equation}
	Using the first-order result in Eq.~\eqref{eq:rho1_vg}, with
	\((b,\omega_1)\) replaced by \((c,\omega_2)\), Eq.~\eqref{eq:v_rho1_comm_vg}
	gives
	\begin{equation}
		\begin{split}
			\comm{v^b}{\rho^{(1)c}}_{mn}
			=
			-\frac{q}{\hbar}A^c
			\sum_{\ell}
			\left[
			\frac{f_{\ell n}v^b_{m\ell}v^c_{\ell n}}
			{\omega_{\ell n}-\omega_2}
			-
			\frac{f_{m\ell}v^c_{m\ell}v^b_{\ell n}}
			{\omega_{m\ell}-\omega_2}
			\right].
		\end{split}
		\label{eq:v_rho1_comm_eval_vg}
	\end{equation}
	Using Eqs.~\eqref{eq:v_rho1_comm_eval_vg} and
	\eqref{eq:w_rho0_comm_vg} in Eq.~\eqref{eq:rho2_vg_band}, and separating
	the endpoint terms \(\ell=m\) and \(\ell=n\) in the internal band sum, we
	obtain the off-diagonal second-order density matrix, for \(m\neq n\),
	\begin{equation}
		\begin{split}
			\rho^{(2)}_{mn}
			&=
			\frac{-q^2A^bA^c}
			{\hbar(\omega_{mn}-\Omega)}
			\Bigg[
			\sum_{\ell\neq m,n}
			\left(
			\frac{f_{\ell n}v^b_{m\ell}v^c_{\ell n}}
			{\epsilon_{\ell n}-\hbar\omega_2}
			-
			\frac{f_{m\ell}v^c_{m\ell}v^b_{\ell n}}
			{\epsilon_{m\ell}-\hbar\omega_2}
			\right)\\
			&+\frac{f_{mn}\Delta^b_{mn}v^c_{mn}}
			{\epsilon_{mn}-\hbar\omega_2}
			+\frac{1}{2}f_{nm}w^{bc}_{mn}
			+(b,\omega_1)\leftrightarrow(c,\omega_2)
			\Bigg],
		\end{split}
		\label{eq:rho2_vg_ordered}
	\end{equation}
	where
	\begin{equation}
		\Delta^b_{mn}
		\equiv
		v^b_{mm}-v^b_{nn}.
	\end{equation}
	
	Unlike the first-order density matrix, the second-order density matrix has a
	nonzero diagonal component in the velocity gauge. Setting \(m=n\) in
	Eq.~\eqref{eq:rho2_vg_band}, the \(w^{bc}\) term vanishes because
	\([w^{bc},\rho^{(0)}]_{nn}=0\). Thus the diagonal component is generated
	entirely by the commutator between \(v^b\) and the first-order interband
	density matrix:
	\begin{equation}
		-i\Omega\rho^{(2)}_{nn}
		=
		\frac{iq}{\hbar}A^b
		\comm{v^b}{\rho^{(1)c}}_{nn}
		+(b,\omega_1)\leftrightarrow(c,\omega_2).
	\end{equation}
	Using Eq.~\eqref{eq:v_rho1_comm_eval_vg}, we obtain
	\begin{equation}
		\begin{split}
			\rho^{(2)}_{nn}
			&=
			\frac{q^2A^bA^c}{\hbar\Omega}
			\sum_{m\neq n}
			\left[
			\frac{f_{mn}v^b_{nm}v^c_{mn}}
			{\epsilon_{mn}-\hbar\omega_2}
			-
			\frac{f_{nm}v^c_{nm}v^b_{mn}}
			{\epsilon_{nm}-\hbar\omega_2}
			\right]  \\
			&\quad
			+(b,\omega_1)\leftrightarrow(c,\omega_2).
		\end{split}
		\label{eq:rho2_vg_diag}
	\end{equation}
	
	\section{Matrix-element algebra in the velocity gauge}
	\subsection{Matrix elements of \(w^{ab}\)}
	\label{app:wab_matrix_element}
	
	In the velocity gauge derivation, we need the matrix element
	\begin{equation}
		w^{ab}_{mn}
		\equiv
		\frac{1}{\hbar^2}\mel{m}{\partial_a\partial_b H_0}{n}.
	\end{equation}
	It is useful to first derive a general identity for a band-space operator
	\(O\). Differentiating \(O_{mn}=\mel{m}{O}{n}\), we have
	\begin{equation}
		\partial_a O_{mn}
		=
		\mel{\partial_a m}{O}{n}
		+
		\mel{m}{\partial_a O}{n}
		+
		\mel{m}{O}{\partial_a n}.
	\end{equation}
	Inserting the identity \(\sum_\ell\ket{\ell}\bra{\ell}=\mathds 1\) and using
	Eq.~\eqref{eq:diagA_overlap}, we obtain
	\begin{equation}
		\mel{m}{\partial_a O}{n}
		=
		\partial_a O_{mn}
		-
		i\sum_\ell
		\left(
		\mathcal{A}^a_{m\ell}O_{\ell n}
		-
		O_{m\ell}\mathcal{A}^a_{\ell n}
		\right).
		\label{eq:operator_derivative_identity}
	\end{equation}
	
	Taking \(O=\hbar^{-1}\partial_b H_0\), so that \(O_{mn}=v^b_{mn}\), gives
	\begin{equation}
		\hbar w^{ab}_{mn}
		=
		\partial_a v^b_{mn}
		-
		i\sum_\ell
		\left(
		\mathcal{A}^a_{m\ell}v^b_{\ell n}
		-
		v^b_{m\ell}\mathcal{A}^a_{\ell n}
		\right).
		\label{eq:wab_from_operator_identity}
	\end{equation}
	Using
	\begin{equation}
		v^b_{mn}
		=
		i\omega_{mn}r^b_{mn}
		+\delta_{mn}\partial_b\omega_m,
		\label{eq:vb_matrix_element}
	\end{equation}
	we obtain
	\begin{equation}
		\begin{split}
			\hbar w^{ab}_{mn}
			&=
			i\Delta^a_{mn}r^b_{mn}
			+i\omega_{mn}\partial_a r^b_{mn}
			+\delta_{mn}\partial_a\partial_b\omega_m \\
			&\quad
			+i\Delta^b_{mn}\mathcal{A}^a_{mn}
			+\sum_\ell
			\left(
			\omega_{\ell n}\mathcal{A}^a_{m\ell}r^b_{\ell n}
			-
			\omega_{m\ell}r^b_{m\ell}\mathcal{A}^a_{\ell n}
			\right).
		\end{split}
		\label{eq:wab_intermediate}
	\end{equation}
	
	We now separate the \(\ell=m\), \(\ell=n\), and
	\(\ell\neq m,n\) terms in the last sum. This makes the covariant structure
	explicit: the diagonal Berry connections combine with
	\(\partial_a r^b_{mn}\) to form the generalized derivative
	Eq.~\eqref{eq:r_gen_deriv}, while the remaining terms are purely interband
	and can be compared directly with the length gauge result. For off-diagonal
	factors, \(\mathcal{A}^a_{mn}=r^a_{mn}\), while the diagonal Berry
	connections \(\mathcal{A}^a_{mm}\) and \(\mathcal{A}^a_{nn}\) are kept
	explicitly. This gives
	\begin{equation}
		\begin{split}
			\hbar w^{ab}_{mn}
			&=
			i\Delta^a_{mn}r^b_{mn}
			+i\Delta^b_{mn}r^a_{mn}
			+\delta_{mn}\partial_a\partial_b\omega_m\\
			&\quad
			+i\omega_{mn}\partial_a r^b_{mn}
			+\omega_{mn}
			\left(\mathcal{A}^a_{mm}-\mathcal{A}^a_{nn}\right)r^b_{mn}\\
			&\quad
			+\sum_{\ell\neq m,n}
			\left(
			\omega_{\ell n}r^a_{m\ell}r^b_{\ell n}
			-
			\omega_{m\ell}r^b_{m\ell}r^a_{\ell n}
			\right)\\
			&=
			i\Delta^a_{mn}r^b_{mn}
			+i\Delta^b_{mn}r^a_{mn}
			+\delta_{mn}\partial_a\partial_b\omega_m
			+i\omega_{mn}r^{b;a}_{mn}\\
			&\quad
			+\sum_{\ell\neq m,n}
			\left(
			\omega_{\ell n}r^a_{m\ell}r^b_{\ell n}
			-
			\omega_{m\ell}r^b_{m\ell}r^a_{\ell n}
			\right).
		\end{split}
		\label{eq:wab_final}
	\end{equation}
	For later use in the current response, we also record the diagonal component.
	Setting \(m=n\) in Eq.~\eqref{eq:wab_final} gives
	\begin{equation}
		\hbar w^{ab}_{nn}
		=
		\partial_a\partial_b\omega_n
		+
		\sum_{\ell\neq n}
		\left(
		\omega_{\ell n}r^a_{n\ell}r^b_{\ell n}
		-
		\omega_{n\ell}r^b_{n\ell}r^a_{\ell n}
		\right).
		\label{eq:wab_diag}
	\end{equation}
	Equivalently, using \(\omega_{n\ell}=-\omega_{\ell n}\),
	\begin{equation}
		\hbar w^{ab}_{nn}
		=
		\partial_a\partial_b\omega_n
		+
		\sum_{\ell\neq n}
		\omega_{\ell n}
		\left(
		r^a_{n\ell}r^b_{\ell n}
		+
		r^b_{n\ell}r^a_{\ell n}
		\right).
		\label{eq:wab_diag_sym}
	\end{equation}

	\subsection{Matrix element of \(u^{abc}\)}
	
	The same operator-derivative identity also gives the third-order velocity
	vertex. By definition,
	\begin{equation}
		u^{abc}_{mn}
		\equiv
		\hbar^{-3}\bra{m}\partial_a\partial_b\partial_c H_0\ket{n}
		=
		\hbar^{-1}\bra{m}\partial_a w^{bc}\ket{n}.
	\end{equation}
	Taking \(O=w^{bc}\) in Eq.~\eqref{eq:operator_derivative_identity}, we obtain
	\begin{equation}
		\hbar u^{abc}_{mn}
		=
		\partial_a w^{bc}_{mn}
		-
		i\sum_\ell
		\left(
		\mathcal{A}^a_{m\ell}w^{bc}_{\ell n}
		-
		w^{bc}_{m\ell}\mathcal{A}^a_{\ell n}
		\right).
		\label{eq:uabc_general}
	\end{equation}
	Separating the \(\ell=m\), \(\ell=n\), and \(\ell\neq m,n\) terms gives
	\begin{equation}
		\begin{split}
			\hbar u^{abc}_{mn}
			&=
			\partial_a w^{bc}_{mn}
			-i\left(\mathcal{A}^a_{mm}-\mathcal{A}^a_{nn}\right)w^{bc}_{mn}\\
			&\quad+i (w^{bc}_{mm}-w^{bc}_{nn})\mathcal{A}^a_{mn} \\
			&\quad
			-i\sum_{\ell\neq m,n}
			\left(
			r^a_{m\ell}w^{bc}_{\ell n}
			-
			w^{bc}_{m\ell}r^a_{\ell n}
			\right).
		\end{split}
		\label{eq:uabc_split}
	\end{equation}
	Equivalently, defining the generalized derivative of \(w^{bc}_{mn}\) as
	\begin{equation}
		w^{bc;a}_{mn}
		\equiv
		\partial_a w^{bc}_{mn}
		-i\left(\mathcal{A}^a_{mm}-\mathcal{A}^a_{nn}\right)w^{bc}_{mn},
	\end{equation}
	we can write
	\begin{equation}
		\begin{split}
			\hbar u^{abc}_{mn}
			=
			w^{bc;a}_{mn}+i (w^{bc}_{mm}-w^{bc}_{nn})\mathcal{A}^a_{mn} 
			\\-i\sum_{\ell\neq m,n}
			\left(
			r^a_{m\ell}w^{bc}_{\ell n}
			-
			w^{bc}_{m\ell}r^a_{\ell n}
			\right).   
		\end{split}
		\label{eq:uabc_final}
	\end{equation}
	
	For the diagonal component needed in the second-order conductivity tensor,
	the diagonal Berry-connection terms cancel. Therefore,
	\begin{equation}
		\hbar u^{abc}_{nn}
		=
		\partial_a w^{bc}_{nn}
		-i\sum_{m\neq n}
		\left(
		r^a_{nm}w^{bc}_{mn}
		-
		w^{bc}_{nm}r^a_{mn}
		\right).
		\label{eq:uabc_diag_raw}
	\end{equation}
	Using Eq.~\eqref{eq:wab_diag_sym}, we obtain
	\begin{equation}
		\begin{split}
			\partial_a w^{bc}_{nn}
			&=
			\frac{1}{\hbar}
			\Bigg[
			\partial_a\partial_b\partial_c\omega_n
			+
			\sum_{m\neq n}\Delta^a_{mn}\left(r^b_{nm}r^c_{mn}+r^c_{nm}r^b_{mn}\right) \\
			+
			\sum_{m\neq n}&
			\omega_{mn}
			\left(r^{b;a}_{nm}r^c_{mn}+ r^b_{nm}r^{c;a}_{mn}+r^{c;a}_{nm}r^b_{mn}+r^c_{nm}r^{b;a}_{mn}\right)
			\Bigg],
		\end{split}
		\label{eq:partiala_wbc_diag}
	\end{equation}
	where
	\(\Delta^a_{mn}=\partial_a\omega_m-\partial_a\omega_n\). Substituting this
	into Eq.~\eqref{eq:uabc_diag_raw} gives
	\begin{equation}
		\begin{split}
			&\hbar u^{abc}_{nn}
			=
			\frac{1}{\hbar}
			\Bigg[
			\partial_a\partial_b\partial_c\omega_n
			+\sum_{m\neq n}\Delta^a_{mn}
			\left(r^b_{nm}r^c_{mn}+r^c_{nm}r^b_{mn}\right) \\
			&+\sum_{m\neq n}\omega_{mn}
			\left(r^{b;a}_{nm}r^c_{mn}+r^b_{nm}r^{c;a}_{mn}+r^{c;a}_{nm}r^b_{mn}+r^c_{nm}r^{b;a}_{mn}\right)
			\Bigg] \\
			&-i\sum_{m\neq n}
			\left(r^a_{nm}w^{bc}_{mn}-w^{bc}_{nm}r^a_{mn}\right).
		\end{split}
		\label{eq:uabc_diag}
	\end{equation}

	\section{Conductivity tensor in the velocity gauge}
	\label{app:conductivity_velocity_gauge}
	
	We now use the velocity gauge current operators defined in the main text,
	\begin{equation}
		j^{(0)a}=qv^a,\qquad
		j^{(1)a}(\omega_1)=-q^2A^b(\omega_1)w^{ab},
		\label{eq:J1}
	\end{equation}
	and
	\begin{equation}
		j^{(2)a}(\omega_1,\omega_2)
		=
		\frac{q^3}{2}A^b(\omega_1)A^c(\omega_2)u^{abc}
		+(b,\omega_1)\leftrightarrow(c,\omega_2).
		\label{eq:j2}
	\end{equation}
	Together with the density matrices derived above, these operators determine
	the linear and second-order conductivity tensors.
	
	\subsection{Linear order response}
	\label{appendix:linear_response_vg}
	
	At linear order, the velocity gauge current receives contributions from both
	the ordinary current operator \(j^{(0)a}\rho^{(1)}\) and the field-dependent
	current operator \(j^{(1)a}\rho^{(0)}\):
	\begin{equation}
		J^{a,(1)}(\omega)
		=
		\Tr\left[
		j^{(0)a}\rho^{(1)}(\omega)
		+
		j^{(1)a}(\omega)\rho^{(0)}
		\right].
		\label{eq:J1_vg_trace}
	\end{equation}
	Using Eq.~\eqref{eq:J1} and Eq.~\eqref{eq:rho1_vg}, we obtain
	\begin{equation}
		\begin{split}
			J^{a,(1)}
			&=
			-q^2A^b\sum_{m\neq n}
			\frac{f_{mn}v^a_{nm}v^b_{mn}}{\epsilon_{mn}-\hbar\omega}
			-q^2A^b\sum_n f_n w^{ab}_{nn}.
		\end{split}
		\label{eq:J1_vg_after_rho1}
	\end{equation}
	Using \(E^b=i\omega A^b\), we define
	\(J^{a,(1)}=\sigma^{ab}_{\rm VG}(\omega)E^b\). Dividing
	Eq.~\eqref{eq:J1_vg_after_rho1} by \(E^b\) gives the velocity gauge
	linear conductivity
	\begin{equation}
		\begin{split}
			\sigma^{ab}_{\rm VG}(\omega)
			&=
			\frac{i q^2}{\omega}
			\sum_{m\neq n}
			\frac{f_{mn}v^a_{nm}v^b_{mn}}{\epsilon_{mn}-\hbar\omega}
			+
			\frac{i q^2}{\omega}
			\sum_n f_n w^{ab}_{nn}.
		\end{split}
		\label{eq:sigma_vg_raw}
	\end{equation}
	
	To compare with the length gauge result, we use
	Eqs.~\eqref{eq:vb_matrix_element} and \eqref{eq:wab_diag_sym}. The
	velocity gauge conductivity becomes
	\begin{equation}
		\begin{split}
			\sigma^{ab}_{\rm VG}(\omega)
			&=
			\frac{i q^2}{\hbar \omega}
			\Bigg[
			\sum_{m\neq n}
			\frac{f_{mn}\omega_{mn}^2 r^a_{nm}r^b_{mn}}
			{\omega_{mn}-\omega}
			+
			\sum_n f_n\partial_a\partial_b\omega_n  \\
			&\qquad
			+
			\sum_{m\neq n}
			f_n\omega_{mn}
			\left(
			r^a_{nm}r^b_{mn}
			+
			r^b_{nm}r^a_{mn}
			\right)
			\Bigg].
		\end{split}
		\label{eq:sigma_vg_raw_r}
	\end{equation}
	The second term can be integrated by parts,
	\begin{equation}
		\sum_n f_n\partial_a\partial_b\omega_n
		=
		-\sum_n v^a_{nn}\partial_b f_n .
	\end{equation}
	For the interband terms, interchanging \(m\leftrightarrow n\) in the last
	term where needed gives
	\begin{equation}
		\begin{split}
			&\sum_{m\neq n}
			\frac{f_{mn}\omega_{mn}^2 r^a_{nm}r^b_{mn}}
			{\omega_{mn}-\omega}
			+
			\sum_{m\neq n}
			f_n\omega_{mn}
			\left(
			r^a_{nm}r^b_{mn}
			+
			r^b_{nm}r^a_{mn}
			\right) \\
			&\qquad
			=
			\sum_{m\neq n}
			\left[
			\frac{f_{mn}\omega_{mn}^2}
			{\omega_{mn}-\omega}
			-
			f_{mn}\omega_{mn}
			\right]
			r^a_{nm}r^b_{mn} \\
			&\qquad
			=
			\omega
			\sum_{m\neq n}
			\frac{f_{mn}\omega_{mn} r^a_{nm}r^b_{mn}}
			{\omega_{mn}-\omega}.
		\end{split}
		\label{eq:sigma_vg_interband_simplify}
	\end{equation}
	Therefore,
	\begin{equation}
		\begin{split}
			\sigma^{ab}_{\rm VG}(\omega)
			&=
			-\frac{q^2}{\hbar}
			\sum_n v^a_{nn}\frac{i\partial_b f_n}{\omega}
			+
			\frac{i q^2}{\hbar}
			\sum_{m\neq n}
			\frac{f_{mn}\omega_{mn}r^a_{nm}r^b_{mn}}
			{\omega_{mn}-\omega}.
		\end{split}
		\label{eq:sigma_vg_final_r}
	\end{equation}
	This agrees exactly with the length gauge result in Eq.~\eqref{eq:sigma_ab}
	using \(v^a_{nm}=i\omega_{nm}r^a_{nm}\).
	
	Since Eq.~\eqref{eq:sigma_vg_final_r} is identical to the length gauge
	conductivity in Eq.~\eqref{eq:sigma_ab}, all frequency limits, including the
	adiabatic dc prescription \(\omega\to i\eta\) with \(\eta\to0^+\), follow
	directly from the length gauge result. This explicitly demonstrates the equivalence between the velocity gauge and length gauge density-matrix formulations at linear order, as required by
	gauge invariance.
	\subsection{second-order response}
	\label{app:vg_2ndorder}
	At second-order, the velocity gauge current contains three distinct
	contributions,
	\begin{equation}
		J^{a,(2)}(\Omega;\omega_1,\omega_2)
		=
		J_0^a+J_1^a+J_2^a,
		\label{eq:J2_vg_decomp}
	\end{equation}
	with
	\begin{equation}
		J_0^a=
		\Tr\left[j^{a,(0)}\rho^{(2)}\right],
		\qquad
		J_1^a=
		\Tr\left[j^{a,(1)}(\omega_1)\rho^{(1)c}(\omega_2)\right],
	\end{equation}
	and
	\begin{equation}
		J_2^a=
		\Tr\left[j^{a,(2)}(\omega_1,\omega_2)\rho^{(0)}\right],
		\label{eq:J2a}
	\end{equation}
	together with the interchanged contribution
	\((b,\omega_1)\leftrightarrow(c,\omega_2)\). This decomposition has no
	direct analogue in the length gauge, where the current operator is field
	independent. The equivalence is recovered only after all three terms are
	combined.
	
	\subsubsection{\texorpdfstring{\(J_0\): contribution from \(j^{(0)}\rho^{(2)}\)}{J0 contribution from j0 rho2}}
	
	We first consider the contribution from the field-independent current operator
	\(j^{(0)a}=qv^a\). Since the second-order density matrix contains both
	off-diagonal and diagonal components in the velocity gauge, we separate
	\begin{equation}
		J_0^a
		=
		q\sum_{m\neq n}v^a_{nm}\rho^{(2)}_{mn}
		+
		q\sum_n v^a_{nn}\rho^{(2)}_{nn}
		\equiv
		J^a_{0,\mathrm{od}}+J^a_{0,\mathrm{diag}} .
		\label{eq:J0_vg_split}
	\end{equation}
	Here and below we suppress the Brillouin-zone integral and write
	\(A^b\equiv A^b(\omega_1)\), \(A^c\equiv A^c(\omega_2)\), and
	\(\Omega=\omega_1+\omega_2\). The contribution with
	\((b,\omega_1)\leftrightarrow(c,\omega_2)\) is understood. Since
	\(E^b(\omega_1)E^c(\omega_2)=-\omega_1\omega_2 A^bA^c\), the corresponding
	conductivity contribution is
	\begin{equation}
		\chi^{abc}_{0}
		=
		-\frac{J^a_{0}}{\omega_1\omega_2 A^bA^c}.
		\label{eq:chi0_from_J0_vg}
	\end{equation}
	
	Using the off-diagonal second-order density matrix in
	Eq.~\eqref{eq:rho2_vg_ordered}, we obtain ~\cite{Moore2019}
	\begin{equation}
		\begin{split}
			\chi^{abc}_{0,\mathrm{od}}
			&=
			\frac{q^3}{\omega_1\omega_2}
			\sum_{n,m\neq n}
			\frac{v^a_{nm}}
			{\hbar(\omega_{mn}-\Omega)}
			\Bigg[
			\frac{f_{mn}\Delta^b_{mn}v^c_{mn}}
			{\epsilon_{mn}-\hbar\omega_2}\\
			&-\frac{1}{2}f_{mn}w^{bc}_{mn} 
			+\sum_{\ell\neq m,n}
			\left(
			\frac{f_{\ell n}v^b_{m\ell}v^c_{\ell n}}
			{\epsilon_{\ell n}-\hbar\omega_2}
			-
			\frac{f_{m\ell}v^c_{m\ell}v^b_{\ell n}}
			{\epsilon_{m\ell}-\hbar\omega_2}
			\right) \\
			&
			+(b,\omega_1)\leftrightarrow(c,\omega_2)
			\Bigg].
		\end{split}
		\label{eq:chi0_off_vg}
	\end{equation}
	
	Using the diagonal second-order density matrix in
	Eq.~\eqref{eq:rho2_vg_diag}, we obtain
	\begin{equation}
		\begin{split}
			\chi^{abc}_{0,\mathrm{diag}}
			&=
			-\frac{q^3}{\hbar^2\Omega\,\omega_1\omega_2}
			\sum_{n,m\neq n}
			v^a_{nn}
			\Bigg[
			\frac{f_{mn}v^b_{nm}v^c_{mn}}
			{\omega_{mn}-\omega_2}\\
			&-\frac{f_{nm}v^c_{nm}v^b_{mn}}
			{\omega_{nm}-\omega_2}
			\Bigg]  
			+(b,\omega_1)\leftrightarrow(c,\omega_2).
		\end{split}
		\label{eq:chi0_diag_vg_raw}
	\end{equation}
	Relabeling \(m\leftrightarrow n\) in the second term gives
	\begin{equation}
		\begin{split}
			\chi^{abc}_{0,\mathrm{diag}}
			&=
			\frac{q^3}{\hbar^2\Omega\,\omega_1\omega_2}
			\sum_{n,m\neq n}
			\frac{f_{mn}\Delta^a_{mn}v^b_{nm}v^c_{mn}}
			{\omega_{mn}-\omega_2}\\
			&+(b,\omega_1)\leftrightarrow(c,\omega_2).
		\end{split}
		\label{eq:chi0_diag_vg}
	\end{equation}
	
	The two terms above are the \(j^{(0)}\rho^{(2)}\) contribution to the
	velocity gauge conductivity. They are not separately gauge invariant and must
	be combined with the field-dependent-current contributions
	\(\chi^{abc}_{1}\) and \(\chi^{abc}_{2}\), derived below, before comparison
	with the length gauge result.

	\subsubsection{Contributions from field-dependent current operators}
	
	We next consider the contributions from the field-dependent current operators,
	\(j^{a,(1)}\rho^{(1)}\) and \(j^{a,(2)}\rho^{(0)}\). For the ordered
	contribution proportional to \(A^b(\omega_1)A^c(\omega_2)\),
	\begin{equation}
		J_1^a
		=
		\Tr\left[
		j^{a,(1)}(\omega_1)\rho^{(1)c}(\omega_2)
		\right],
		\quad
		j^{a,(1)}(\omega_1)=-q^2A^b w^{ba}.
	\end{equation}
	Since \(\rho^{(1)}_{nn}=0\), only off-diagonal matrix elements contribute.
	Although \(w^{ab}=w^{ba}\) as an operator, we write the current operator as \(w^{ba}\) here because its band-basis decomposition naturally contains \(r^{a;b}_{mn}\), matching the generalized derivative structure used in the length gauge expression.
	Using Eq.~\eqref{eq:rho1_vg}, with
	\((b,\omega_1)\rightarrow(c,\omega_2)\), and dividing by
	\(E^b(\omega_1)E^c(\omega_2)=-\omega_1\omega_2A^bA^c\), we obtain
	\begin{equation}
		\chi^{abc}_{1}
		=
		-\frac{q^3}{\omega_1\omega_2}
		\sum_{m\neq n}
		\frac{f_{mn}w^{ba}_{nm}v^c_{mn}}
		{\epsilon_{mn}-\hbar\omega_2}
		+(b,\omega_1)\leftrightarrow(c,\omega_2).
		\label{eq:chi1_vg}
	\end{equation}

	The final field-dependent-current contribution comes from
	\(j^{(2)a}\rho^{(0)}\). Since the equilibrium density matrix is diagonal,
	\(\rho^{(0)}_{mn}=f_n\delta_{mn}\), only the diagonal matrix element of the
	second-order current operator contributes. Using Eq.~\eqref{eq:j2} in
	Eq.~\eqref{eq:J2a} and converting \(A^bA^c\) to \(E^bE^c\), we obtain
	\begin{equation}
		\begin{split}
			\chi^{abc}_{2}
			&=
			-\frac{q^3}{2\omega_1\omega_2}
			\sum_n f_n u^{abc}_{nn}
			+
			(b,\omega_1)\leftrightarrow(c,\omega_2).
		\end{split}
		\label{eq:chi2_vg_start}
	\end{equation}
	Although this contribution is diagonal in the equilibrium density matrix, it
	is not purely intraband. As shown in Eq.~\eqref{eq:uabc_diag}, the diagonal
	matrix element \(u^{abc}_{nn}\) contains both the derivative
	\(\partial_a w^{bc}_{nn}\) and the commutator-like interband contribution
	involving \(w^{bc}_{mn}\). Thus \(\chi^{abc}_{2}\) must be kept together with
	\(\chi^{abc}_{0,\mathrm{od}}\), \(\chi^{abc}_{0,\mathrm{diag}}\), and
	\(\chi^{abc}_{1}\) when taking the dc adiabatic limit.

	\subsubsection{Static limit and reduction to the length gauge form}
	The full velocity gauge response is obtained from the sum
	\begin{equation}
		\chi^{abc}_{\rm VG}
		=
		\chi^{abc}_{0,\mathrm{od}}
		+
		\chi^{abc}_{0,\mathrm{diag}}
		+
		\chi^{abc}_{1}
		+
		\chi^{abc}_{2}.
	\end{equation}
	This full expression is expected to be equivalent to the length gauge
	conductivity at finite frequency, provided all field-dependent current
	operators are retained. However, this equivalence is not manifest term by
	term. In the velocity gauge, the decomposition into
	\(j^{(0)}\rho^{(2)}\), \(j^{(1)}\rho^{(1)}\), and \(j^{(2)}\rho^{(0)}\) does
	not map directly onto the diagonal and off-diagonal sectors of the
	length gauge calculation. In particular, the field-dependent current
	operators contain interband geometric matrix elements, so the two-band and
	three-band structures are redistributed among the four terms above.
	
	We therefore focus on the same physical adiabatic dc trajectory [Eq.~\eqref{eq:dc_prescription_second}] used in the
	length gauge analysis. Rather than rearranging the full finite frequency
	expression into the length gauge form, we expand the complete velocity gauge
	response along this trajectory and extract the \(\eta^0\) term. This finite
	term can then be compared directly with the length gauge dc result.
	
	Along the static adiabatic trajectory
	\(\omega_1=\omega_2=i\eta\), \(\Omega=2i\eta\), we expand the complete
	velocity gauge response in powers of \(\eta\) and retain the finite
	\(\eta^0\) coefficient. Suppressing the common prefactor \(q^3/\hbar^2\), we obtain
	\begin{widetext}
		\begin{equation}
			\begin{split}
				\chi^{abc}_{\rm VG}&=
				\sum_{\substack{n,m \neq n \\ \ell \neq m,n}}
				\left[
				f_{\ell n}v^a_{nm}v^b_{m\ell}v^c_{\ell n}
				\left(
				\frac{1}{\omega_{mn}\omega_{\ell n}^3}
				+\frac{2}{\omega_{mn}^2\omega_{\ell n}^2}
				+\frac{4}{\omega_{mn}^3\omega_{\ell n}}
				\right)-
				f_{m\ell}v^a_{nm}v^c_{m\ell}v^b_{\ell n}
				\left(\frac{1}{\omega_{mn}\omega_{m\ell}^3}+\frac{2}{\omega_{mn}^2\omega_{m\ell}^2}+\frac{4}{\omega_{mn}^3\omega_{m\ell}}\right)\right]\\
				&+\sum_{n,m\neq n}f_{mn}\frac{7v^a_{nm}\Delta^b_{mn}v^c_{mn}}
				{\omega_{mn}^4}
				-2f_{mn}\frac{v^a_{nm}\hbar w^{bc}_{mn}}{\omega_{mn}^3}
				+f_{mn}\frac{\Delta^a_{mn}v^b_{nm}v^c_{mn}}
				{2\omega_{mn}^4}
				-f_{mn}\frac{\hbar w^{ba}_{nm}v^c_{mn}}
				{\omega^3_{mn}}
				+
				(b\leftrightarrow c).
			\end{split}
			\label{eq:chi_static_vg_eta0}
		\end{equation}
	\end{widetext}
	Note that the \(\chi_2\) contribution does not generate a separate
	\(\eta^0\) term in this expansion; it contributes only to the singular
	Drude-like vector potential response. Equation~\eqref{eq:chi_static_vg_eta0} can be further
	organized by separating terms according to whether they involve an intermediate
	third band. In doing so, one must also expand the \(w\)-matrix elements,
	because \(w^{ab}_{mn}\) contains both two-band generalized derivative terms
	and three-band products. For example, one of the \(w\)-dependent terms has the structure
	\begin{equation}
		\sum_{n,m\neq n}
		2if_{mn}\frac{r^a_{nm}\hbar w^{bc}_{mn}}{\omega_{mn}^2}.
	\end{equation}
	Using Eq.~\eqref{eq:wab_final}, this separates into a two-band part and a
	three-band part:
	\begin{equation}
		\begin{split}
			&
			\sum_{n,m\neq n}
			-2f_{mn}\frac{r^a_{nm}}{\omega_{mn}^2}
			\Big(
			\Delta^b_{mn}r^c_{mn}
			+\Delta^c_{mn}r^b_{mn}
			+\omega_{mn}r^{c;b}_{mn}
			\Big) \\
			&+2i
			\sum_{\substack{n,m\neq n\\ \ell\neq m,n}}
			f_{mn}\frac{r^a_{nm}}{\omega_{mn}^2}
			\left(
			\omega_{\ell n}r^b_{m\ell}r^c_{\ell n}
			-
			\omega_{m\ell}r^c_{m\ell}r^b_{\ell n}
			\right).
		\end{split}
		\label{eq:rw_example_split}
	\end{equation}
	
	\paragraph{Simplification of triple-\(r\) terms.}
	
	We now collect the triple-\(r\) terms. To keep the expressions compact, we
	use a Fermi-weighted commutator notation
	\cite{Kaplan2023,Yan2023}. For two off-diagonal band matrices \(A\) and
	\(B\), we define
	\begin{equation}
		\begin{split}
			[A,B]_f
			&\equiv
			\sum_{n,m} f_n
			\left(
			A_{nm}B_{mn}-B_{nm}A_{mn}
			\right)\\
			&=
			\sum_{n,m} f_{nm}A_{nm}B_{mn}.
		\end{split}
		\label{eq:fermi_weighted_comm}
	\end{equation}
	The second equality follows by relabeling dummy band indices.
	
	For example, one of the triple-\(r\) structures generated by the
	\(w^{bc}_{mn}\) term in Eq.~\eqref{eq:rw_example_split} can be written as
	\begin{equation}
		\begin{split}
			&[\frac{r^a}{\omega^2},[r^b,\omega r^c]]_f
			=
			\sum_{n,m}
			f_{nm}\frac{r^a_{nm}}{\omega_{nm}^2}
			[r^b,\omega r^c]_{mn} \\
			&=
			\sum_{n,m,\ell}
			f_{nm}\frac{r^a_{nm}}{\omega_{nm}^2}
			\left(
			r^b_{m\ell}\omega_{\ell n}r^c_{\ell n}
			-
			\omega_{m\ell}r^c_{m\ell}r^b_{\ell n}
			\right).
		\end{split}
		\label{eq:compact_comm_example}
	\end{equation}
	In this triple-\(r\) collection, \(r^a\) denotes the interband position
	matrix element, so \(r^a_{nn}=0\). Thus products such as
	\(r^a_{nm}r^b_{m\ell}r^c_{\ell n}\) automatically restrict the internal
	indices to distinct bands, \(m\neq n\), \(\ell\neq m\), and \(\ell\neq n\).
	The endpoint contributions have already been separated into the two-band
	sector.
	
	With this notation, the triple-\(r\) terms generated by the \(w\)-matrix
	elements are
	\begin{equation}
		T_w^{abc}=
		-2i\,[\frac{r^a}{\omega^2},[r^b,\omega r^c]]_f
		-i\,[\frac{r^c}{\omega^2},[r^b,\omega r^a]]_f+(b\leftrightarrow c).
		\label{eq:Tw_triple_vg}
	\end{equation}
	The explicit three-band terms in Eq.~\eqref{eq:chi_static_vg_eta0} are first brought to a common Fermi-factor
	convention. For the term proportional to \(f_{\ell n}\), we relabel the dummy
	indices cyclically as \((n,m,\ell)\rightarrow(m,\ell,n)\), so that
	\(f_{\ell n}\rightarrow f_{nm}\). For the term proportional to \(f_{m\ell}\),
	we use \((n,m,\ell)\rightarrow(\ell,n,m)\), so that
	\(f_{m\ell}\rightarrow f_{nm}\). After these relabelings, all three-band
	products can be written with the same Fermi factor \(f_{nm}\) and compared
	directly with the Fermi-weighted commutator notation.
	\begin{equation}
		\begin{split}
			T_r^{abc}
			=&
			-4i\,[r^c,[\omega r^b,\frac{r^a}{\omega^2}]]_f
			+2i\,[\frac{r^c}{\omega},[\omega r^b,\frac{r^a}{\omega}]]_f\\
			&
			-i\,[\frac{r^c}{\omega^2},[\omega r^b,r^a]]_f+(b\leftrightarrow c)
		\end{split}
		\label{eq:Tr_triple_vg}
	\end{equation}
	
	The reduction of \(T_r^{abc}\) uses two elementary identities. The first is the
	Jacobi identity,
	\begin{equation}
		[r^c,[\omega r^b,\frac{r^a}{\omega^2}]]_f
		=
		-[\frac{r^a}{\omega^2},[r^c,\omega r^b]]_f
		-[\omega r^b,[\frac{r^a}{\omega^2},r^c]]_f .
		\label{eq:jacobi_triple_1}
	\end{equation}
	The second is the energy-shift identity around a three-band loop: $\omega_{m\ell}= -\omega_{\ell n}-\omega_{nm}$.
	This identity allows factors of \(\omega\) appearing inside the inner
	commutator to be moved onto the outer matrix element. In the compact
	commutator notation this gives, for example,
	\begin{equation}
		\begin{split}
			[r^c,[\omega r^b,\frac{r^a}{\omega^2}]]_f
			&=
			-[r^c,[r^b,\frac{r^a}{\omega}]]_f
			-[\omega r^c,[r^b,\frac{r^a}{\omega^2}]]_f .
		\end{split}
		\label{eq:energy_shift_identity_1}
	\end{equation}
	Similarly,
	\begin{equation}
		\begin{split}
			[\frac{r^c}{\omega},[\omega r^b,\frac{r^a}{\omega}]]_f
			&=
			-[\frac{r^c}{\omega},[r^b,r^a]]_f
			-[r^c,[r^b,\frac{r^a}{\omega}]]_f ,
			\\
			[\frac{r^c}{\omega^2},[\omega r^b,r^a]]_f
			&=
			-[\frac{r^c}{\omega},[r^b,r^a]]_f
			-[\frac{r^c}{\omega^2},[r^b,\omega r^a]]_f .
		\end{split}
		\label{eq:energy_shift_identity_2}
	\end{equation}
	
	Using Eqs.~\eqref{eq:jacobi_triple_1} and
	\eqref{eq:energy_shift_identity_1}, the first term in \(T_r^{abc}\) can be
	rewritten as
	\begin{equation}
		\begin{split}
			-4i\,[r^c,[\omega r^b,\frac{r^a}{\omega^2}]]_f&= 2i\Big([r^c,[r^b,\frac{r^a}{\omega}]]_f
			+[\omega r^c,[r^b,\frac{r^a}{\omega^2}]]_f\\
			&+[\frac{r^a}{\omega^2},[r^c,\omega r^b]]_f
			+[\omega r^b,[\frac{r^a}{\omega^2},r^c]]_f \Big)\\
			&=2i\Big([r^c,[r^b,\frac{r^a}{\omega}]]_f
			+[\frac{r^a}{\omega^2},[r^c,\omega r^b]]_f\Big),
		\end{split}    
	\end{equation}
	where the second and fourth terms cancel after including the
	\((b\leftrightarrow c)\) contribution.
	
	Using Eq.~\eqref{eq:energy_shift_identity_2}, we get
	\begin{equation}
		\begin{split}
			T_r^{abc}
			&=
			2i[r^c,[r^b,\frac{r^a}{\omega}]]_f
			+2i[\frac{r^a}{\omega^2},[r^c,\omega r^b]]_f\\
			&-2i[\frac{r^c}{\omega},[r^b,r^a]]_f
			-2i[r^c,[r^b,\frac{r^a}{\omega}]]_f \\
			&+i[\frac{r^c}{\omega},[r^b,r^a]]_f
			+i[\frac{r^c}{\omega^2},[r^b,\omega r^a]]_f+(b\leftrightarrow c)\\
			&=2i[\frac{r^a}{\omega^2},[r^c,\omega r^b]]_f-i[\frac{r^c}{\omega},[r^b,r^a]]_f\\
			&+i[\frac{r^c}{\omega^2},[r^b,\omega r^a]]_f+(b\leftrightarrow c)
		\end{split}
		\label{eq:Tr_triple_reduced_vg}
	\end{equation}
	
	Combining Eq.~\eqref{eq:Tr_triple_reduced_vg} with the triple-\(r\) terms
	from the \(w\)-matrix elements, Eq.~\eqref{eq:Tw_triple_vg}, the terms with
	\(\omega^{-2}r\) cancel pairwise. The remaining triple-\(r\) sector is
	\begin{equation}
		\begin{split}
			T_{\rm triple}^{abc}
			&=
			i[\frac{r^c}{\omega},[r^a,r^b]]_f+ i[\frac{r^b}{\omega},[r^a,r^c]]_f.
		\end{split}
		\label{eq:triple_vg_final}
	\end{equation}
	This is exactly the same triple-\(r\) structure obtained in the length gauge
	static response, Eq.~\eqref{eq:triple_r_rewrite}.
	
	\paragraph{Simplification of two-band generalized-derivative terms.}
	
	We next simplify the remaining two-band terms in
	Eq.~\eqref{eq:chi_static_vg_eta0}, namely the terms that do not involve an
	intermediate third band. Inserting the two-band part of the \(w\)-matrix
	elements, as illustrated in Eq.~\eqref{eq:rw_example_split}, and using
	\(v^\alpha_{mn}=i\omega_{mn}r^\alpha_{mn}\), we obtain
	\begin{equation}
		\begin{split}
			T^{abc}_{\rm 2b}
			&=
			\sum_{n,m\neq n} f_{mn}
			\Bigg[
			4\frac{\Delta^b_{mn}r^a_{nm}r^c_{mn}}{\omega_{mn}^2}
			-2\frac{\Delta^c_{mn}r^a_{nm}r^b_{mn}}{\omega_{mn}^2} \\
			&
			-\frac{\Delta^a_{mn}r^b_{nm}r^c_{mn}}{2\omega_{mn}^2}
			-\frac{
				2r^a_{nm}r^{c;b}_{mn}
				+
				r^{a;b}_{nm}r^c_{mn}
			}{\omega_{mn}}
			\Bigg]
			+(b\leftrightarrow c).
		\end{split}
		\label{eq:T2b_start_vg}
	\end{equation}
	
	We again use the identity Eq.~\eqref{eq:Delta} to simplify the terms with $\Delta/\omega^2$:
	\begin{equation}
		\begin{split}
			&T^{abc}_{\rm 2b}
			=
			\sum_{n,m} f_{mn}
			\Bigg[
			-4r^a_{nm}r^c_{mn}\partial_b(\omega_{mn}^{-1})
			+2r^a_{nm}r^b_{mn}\partial_c(\omega_{mn}^{-1}) \\
			&
			+\frac{r^b_{nm}r^c_{mn}}{2}\partial_a(\omega_{mn}^{-1})
			-\frac{
				2r^a_{nm}r^{c;b}_{mn}
				+
				r^{a;b}_{nm}r^c_{mn}
			}{\omega_{mn}}
			\Bigg]
			+(b\leftrightarrow c).
		\end{split}
		\label{eq:T2b_start_vg}
	\end{equation}
	
	We now integrate the total derivative terms by parts over the Brillouin zone. This separates the two-band contribution into a Fermi sea part and a Fermi surface part.
	\begin{equation}
		\begin{split}
			T^{abc}_{\rm 2b, sea}
			&=
			\sum_{n,m} F_{mn}
			\Bigg[
			4\partial_b(r^a_{nm}r^c_{mn})-2\partial_c(r^a_{nm}r^b_{mn}) \\
			&
			-\frac{\partial_a(r^b_{nm}r^c_{mn})}{2}
			-2r^a_{nm}r^{c;b}_{mn}-
			r^{a;b}_{nm}r^c_{mn}
			\Bigg]
			+(b\leftrightarrow c);
		\end{split}
		\label{eq:T2b_sea}
	\end{equation}
	\begin{equation}
		\begin{split}
			T^{abc}_{\rm 2b, sur}
			&=\sum_{n,m} \frac{1}{\omega_{mn}}
			\Bigg[
			4r^a_{nm}r^c_{mn}\partial_bf_{mn}
			-2r^a_{nm}r^b_{mn}\partial_cf_{mn}\\
			&-\frac{r^b_{nm}r^c_{mn}}{2}\partial_af_{mn}
			\Bigg]
			+(b\leftrightarrow c);
		\end{split}
		\label{eq:T2b_sur}
	\end{equation}
	where we use $F_{mn}\equiv f_{mn}/\omega_{mn}$ for simplicity. 
	The Fermi sea part can be simplified further using the product rule for generalized derivatives, Eq.~\eqref{eq:genr_prod}, in the same way as in the length gauge derivation. This gives
	\begin{equation}
		\begin{split}
			T^{abc}_{\rm 2b, sea}
			&=
			\sum_{n,m} F_{mn}
			\Bigg[
			2r^a_{nm}r^{c;b}_{mn}+3r^{a;b}_{nm}r^{c}_{mn}-2r^{a;c}_{nm}r^{b}_{mn}\\
			&-2r^a_{nm}r^{b;c}_{mn}
			-\frac{r^{b}_{nm}r^{c;a}_{mn}}{2}-\frac{r^{b;a}_{nm}r^c_{mn}}{2}
			\Bigg]
			+(b\leftrightarrow c)\\
			&=
			\sum_{n,m} F_{mn}
			\Bigg[
			r^c_{mn}(r^{a;b}_{nm}-r^{b;a}_{mn})+
			r^b_{mn}(r^{a;c}_{nm}-r^{c;a}_{mn})\Bigg]
		\end{split}
		\label{eq:T2b_sea_final}
	\end{equation}
	This is the same Fermi-sea structure obtained in the length gauge,
	Eq.~\eqref{eq:sea_combined_commutator_form}. It cancels the triple-\(r\) contribution in Eq.~\eqref{eq:triple_vg_final}.

	The remaining finite contribution is therefore the Fermi surface part. After writing the \((b\leftrightarrow c)\) term explicitly and collecting like terms, Eq.~\eqref{eq:T2b_sur} becomes\begin{equation}
		\begin{split}
			T^{abc}_{\rm 2b, sur}
			&=\sum_{n,m} \frac{1}{\omega_{mn}}
			\Bigg[
			2r^a_{nm}r^c_{mn}\partial_bf_{mn}
			+2r^a_{nm}r^b_{mn}\partial_cf_{mn}\\
			&-r^b_{nm}r^c_{mn}\partial_af_{mn}
			\Bigg]
		\end{split}
		\label{eq:T2b_sur_final}
	\end{equation}
	This is exactly the Fermi surface term obtained in the length gauge static response, Eq.~\eqref{eq:fermi surface length}.
	
	\subsubsection{Alternative \(w^{ab}\) representation and Fermi sea cancellation}\label{app: symmetric wab}
	In the main derivation we choose a particular representation of \(w^{ba}\), so
	that the intermediate Fermi sea terms naturally match the length gauge
	organization. This choice is not unique. Since \(w^{ab}=w^{ba}\) as an
	operator, one may instead use a symmetrized band-basis representation, as in
	Ref.~\cite{Kaplan2023}. In our notation this reads
	\begin{equation}
		\begin{split}
			\hbar w^{ab}_{mn}
			&=
			i\Delta^a_{mn}r^b_{mn}
			+i\Delta^b_{mn}r^a_{mn}
			+\delta_{mn}\partial_a\partial_b\omega_m+\frac{i\omega_{mn}}{2}r^{b;a}_{mn}
			\\
			&+\frac{i\omega_{mn}}{2}r^{a;b}_{mn}
			-\frac{1}{2}[\omega r^b,r^a]-\frac{1}{2}[\omega r^a,r^b].
		\end{split}
		\label{eq:wab_sym}
	\end{equation}
	This form redistributes terms between the two-band generalized-derivative
	sector and the triple-\(r\) sector, but it cannot change the final response.
	
	We now specialize to \(a=y\), \(b=c=x\), following the organization used in
	Ref.~\cite{Kaplan2023}. With the symmetrized representation of \(w^{ab}\) in
	Eq.~\eqref{eq:wab_sym}, our intermediate static expression Eq.~\eqref{eq:chi_static_vg_eta0} takes exactly the
	compact commutator form of Eq.~S15 of Ref.~\cite{Kaplan2023}. The following
	algebra shows that the apparent Fermi sea terms in this representation still
	cancel once the two-band and triple-\(r\) sectors are combined. The two-band sector is
	\begin{equation}
		\begin{split}
			\chi^{yxx}_{\rm 2b}&=\sum_{n,m}f_{nm}\Bigg[-2\frac{r^y_{nm}\Delta^x_{mn}r^x_{mn}}{\omega_{nm}^2}
			+\frac{\Delta^y_{mn}r^x_{nm}r^x_{mn}}{2\omega_{nm}^2}\\
			&-\frac{
				4r^y_{nm}r^{x;x}_{mn}
				+
				r^{y;x}_{nm}r^x_{mn}+r^{x;y}_{nm}r^x_{mn}
			}{2\omega_{nm}}\Bigg],\\
		\end{split}
		\label{eq:chi_yxx_2b_symw}
	\end{equation}
	while the triple-\(r\) sector is
	\begin{equation}
		\begin{split}        
			\chi^{yxx}_{\rm triple}&=-2i[\frac{r^x}{\omega},[r^x, r^y]]_f-\frac{i}{2}[\frac{r^x}{\omega^2},[\omega r^x, r^y]]_f\\
			&+\frac{i}{2}[\frac{r^x}{\omega^2},[\omega r^y, r^x]]_f.
		\end{split}
		\label{eq:chi_yxx_triple_symw}
	\end{equation}
	Here we consider the single ordered \(yxx\) contribution; the exchange
	\(x\leftrightarrow x\) only changes the overall symmetrization convention.
	
	We next separate Eq.~\eqref{eq:chi_yxx_2b_symw} into Fermi sea and
	Fermi surface pieces using Eq.~\eqref{eq:Delta}. The Fermi sea part is
	\begin{equation}
		\begin{split}
			\chi^{yxx}_{\rm 2b, sea}&=\sum_{n,m}F_{nm}\Bigg(2r^{y;x}_{nm}r^x_{mn}+2r^{y}_{nm}r^{x;x}_{mn}
			-r^{x}_{nm}r^{x;y}_{mn}\\
			&-
			2r^y_{nm}r^{x;x}_{mn}
			-
			\frac{1}{2}r^{y;x}_{nm}r^x_{mn}-\frac{1}{2}r^{x;y}_{nm}r^x_{mn}
			\Bigg)\\
			&=\sum_{n,m}\frac{3}{2}F_{nm}r^{x}_{mn}(r^{y;x}_{nm}-r^{x;y}_{nm})\\
			&=\sum_{n,m}-\frac{3i}{2}F_{nm}r^{x}_{mn}[r^y,r^x]_{nm}\\
		\end{split},
		\label{eq:chi_yxx_2b_sea_symw}
	\end{equation}
	where \(F_{nm}\equiv f_{nm}/\omega_{nm}\). In going to the second line, we
	combine terms related by the dummy-index relabeling \(m\leftrightarrow n\).
	In the last line, we use the interband sum rule Eq.~\eqref{eq:AS24}.
	
	The Fermi-surface part is
	\begin{equation}
		\begin{split}
			\chi^{yxx}_{\rm 2b, sur}&=\sum_{n,m}\frac{1}{\omega_{nm}}\left(2r^y_{nm}r^x_{mn}\partial_xf_{nm}-\frac{1}{2}r^x_{nm}r^x_{mn}\partial_yf_{nm}\right)
		\end{split}.
		\label{eq:chi_yxx_2b_FS_symw}
	\end{equation}

	It remains to simplify the triple-\(r\) contribution. The energy-shift identity
	gives
	\begin{equation}
		\begin{split}
			[\frac{r^x}{\omega^2},[\omega r^x, r^y]]_f&=-[\frac{r^x}{\omega},[r^x, r^y]]_f-[\frac{r^x}{\omega^2},[r^x, \omega r^y]]_f\\
			&=-[\frac{r^x}{\omega},[r^x, r^y]]_f+[\frac{r^x}{\omega^2},[\omega r^y,r^x]]_f.
		\end{split}
		\label{eq:symw_triple_identity}
	\end{equation}
	
	Substituting this into Eq.~\eqref{eq:chi_yxx_triple_symw}, we find
	\begin{equation}
		\begin{split}
			\chi^{yxx}_{\rm triple}&=-2i[\frac{r^x}{\omega},[r^x, r^y]]_f-\frac{i}{2}(-[\frac{r^x}{\omega},[r^x, r^y]]_f)\\
			&=\frac{-3i}{2}[\frac{r^x}{\omega},[r^x, r^y]]_f\\
			&=\sum_{n,m}\frac{3i}{2}F_{nm}r^x_{nm}[ r^y,r^x]_{mn}\\
			&=\sum_{n,m}\frac{3i}{2}F_{nm}r^x_{mn}[ r^y,r^x]_{nm}
		\end{split}   
	\end{equation}
	This exactly cancels the Fermi sea term in
	Eq.~\eqref{eq:chi_yxx_2b_sea_symw}. Thus the symmetrized representation of
	\(w^{ab}\) gives the same final result: the only remaining static contribution
	is the Fermi surface term in Eq.~\eqref{eq:chi_yxx_2b_FS_symw}, consistent
	with Eq.~\eqref{eq:T2b_sur_final}.
	\clearpage
	
	\bibliography{references}{}
\end{document}